\documentclass[a4]{article}
\usepackage[T1]{fontenc}
\usepackage[latin1]{inputenc}
\usepackage{amsmath, amssymb, amsthm}
\usepackage[english]{babel}
\usepackage{indentfirst}
\usepackage{booktabs}
\usepackage{array}
\usepackage{caption,appendix}
\usepackage{geometry}
\usepackage{float}
\usepackage{cancel}
\usepackage{bbold}
\usepackage{authblk}
\numberwithin{equation}{section}
\begin{document}
\author[1,2,3]{Salvatore Capozziello}
\author[4]{Maurizio Capriolo} 
\author[4]{Loredana Caso}
\affil[1]{\emph{Dipartimento di Fisica "E. Pancini", Universit\`a di Napoli {}``Federico II'', Compl. Univ. di
		   Monte S. Angelo, Edificio G, Via Cinthia, I-80126, Napoli, Italy, }}
		  \affil[2]{\emph{INFN Sezione  di Napoli, Compl. Univ. di
		   Monte S. Angelo, Edificio G, Via Cinthia, I-80126,  Napoli, Italy,}}
 \affil[3]{\emph{ Tomsk State Pedagogical University, ul. Kievskaya, 60, 634061
Tomsk, Russia, }}
\affil[4]{\emph{Dipartimento di Matematica Universit\`a di Salerno, via Giovanni Paolo II, 132, Fisciano, SA I-84084, Italy.} }
\date{\today}
\title{\textbf{Gravitational Waves in Higher Order Teleparallel Gravity}}
\maketitle
\begin{abstract}
 The teleparallel equivalent of  higher order  Lagrangians like  $L_{\Box R}=-R+a_{0}R^{2}+a_{1}R\Box R$ can be obtained by means of the boundary term $B=2\nabla_{\mu}T^{\mu}$. In this perspective, we  derive the field equations in presence of matter for  higher-order  teleparallel gravity considering, in particular, sixth-order theories where the $\Box$ operator is linearly included. In the weak field approximation,  gravitational wave solutions  for these theories are derived. Three states of polarization are found: the two standard $+$ and $\times$  polarizations, namely 2-helicity massless transverse tensor polarizations, and a 0-helicity massive, with partly transverse and partly longitudinal scalar polarization. Moreover, these gravitational waves exhibit four oscillation modes related to  four degrees of freedom: the two classical $+$ and $\times$ tensor modes of frequency $\omega_{1}$,  related to the standard Einstein waves with $k^{2}_{1}=0$; two mixed longitudinal-transverse scalar modes for each frequencies $\omega_{2}$ and $\omega_{3}$, related to two different  4-wave vectors, $k^{2}_{2}=M_{2}^{2}$ and $k^{2}_{3}=M^{2}_{3}$. The four degrees of freedom are the amplitudes of each individual mode, i.e. $\hat{\epsilon}^{(+)}\left(\omega_{1}\right)$, $\hat{\epsilon}^{(\times)}\left(\omega_{1}\right)$, $\hat{B}_{2}\left(\mathbf{k}\right)$, and $\hat{B}_{3}\left(\mathbf{k}\right)$. To describe a general teleparallel gravity model of  order $(2p + 2)$,  we  used the  teleparallel Lagrangian $L_{\Box^{k}T}^{neq}=e\left(T+a_{0}T^{2}+\sum_{k=1}^{p} a_{k}T\Box^{k}T\right)$ which demonstrates to be not equivalent to $L_{\Box^{k}R}=\sqrt{-g}\left(R+\sum_{k=0}^{p}a_{k}R\Box^{k} R\right)$. By varying its action, the related field equations in the presence of matter are derived. Hence we obtain the gravitational waves for these   teleparallel gravity models, i.e.  $L_{\Box^{k}T}^{neq}$, that are exactly the Einstein gravitational waves as in $f(T)$ teleparallel gravity. In conclusion, the boundary term $B$ generates  extra polarizations and  additional modes beyond the standard ones.  In particular $B^{2}$ and $B\Box B$  terms generate both the additional scalar polarization and the extra scalar modes.  
\end{abstract}



\section{Introduction}
In affine geometry, spacetime can be endowed with a connection whose curvature and torsion do not vanish. Connections are   rules that tell us how to derive  geometrical objects on a manifold and then how to parallel transport them. Furthermore, it is possible to add a metric to the spacetime whose connection is generally non-metric compatible (metric-affine geometry). However, if we want lengths and angles between vectors to be preserved under parallel transport, we have to  impose that the connection is metric-compatible, that is  $\nabla g=0$. After turning off both  curvature and non-metricity in the connection,  gravity appears as a manifestation of the only torsion as  in the Teleparallel Equivalent of General Relativity. In this framework, the autoparallel curves (straight lines), i.e.  the affine geodesics, related to the connection, and the extremal curves (shortest lines), i.e. the metric geodesics or simply the geodesics, related to  metric, are not the same. Autoparallel and extremal curves are the same for  light rays or massless wave traveling at the speed of light, while  the particle trajectories  follow the geodesics as in General Relativity \cite{HHK, FNRM}. Finally, in the Weitzenb\"ock spacetime where  both curvature and non-metricity are zero, we have that the local Lorentz invariance is violated  and the spin connection can vanish. In a nutshell, teleparallelism is a dynamical theory of the vierbein $\left\{e_{a}\right\}$, based on the notion of distance or absolute parallelism $\nabla e_{a}=0$, with respect to the curvature-free  torsion-based connection used in the Weitzenb\"ock spacetime. So gravity is completely described by the torsion two-form $T^{a}=de^{a}$ where the one-forms $\left\{e^{a}\right\}$ are a local basis on the cotangent space \cite{HS}. 

Historically,  teleparallism was proposed by Einstein himself, in 1928, who tried to unify General Relativity with Electromagnetism. The approach was based on replacing the Levi-Civita connection with a non-symmetric connection, the  Weitzenb\"ock connection. The attempt was a failure because the components of the Electromagnetic field, identified with the additional six component of the tetrad, could be eliminated by imposing the local Lorentz invariance.  However, the idea of modifying the spacetime geometry by using the Weitzenb\"ock connection  allowed to formulate a theory of gravity, based on teleparallelism, equivalent to General Relativity, i.e. the Teleparallel Equivalent of General Relativity (TEGR) \cite{MALTEL, CRTG, KSTE}. Teleparallelism and General Relativity describe the same physics because they give the same field equations coming from  the Hilbert-Einstein Lagrangian, linear in the Ricci  curvature scalar $R$ \cite{APTG, CCLS}, and the  Teleparallel Lagrangian, linear  in the scalar torsion $T$. The two Lagrangian  differ by a four-divergence as we will discuss below.

The above theories can be extended in view of Quantum Gravity, considering more general effective Lagragians.
In fact, the issue to renormalize and regularize  General Relativity can be improved  by adding higher-order curvature invariants and related derivatives, minimally or non-minimally coupled with scalar and vector fields in the Hilbert-Einstein action. Improving the actions with higher-order terms  can make the theory more regular and  renormalizable at least at lower loop level \cite{Birrell}. 
Specifically, such contributions appear when the scale lengths approache the Planck scale as one-loop or higher-loop quantum corrections in the effective action. A straightforward extension of the Hilbert-Einstein action can be realized  by assuming  analytical functions of Ricci curvature scalar  $R$, the so called $f(R)$ gravity where   General Relativity is recovered for $f(R)\to R$ \cite{CL,NO,NOO, Petrov}. 

In the same way,  it is possible to extend  TEGR considering  analytical functions $f (T)$  which can be non-linear with respect to $ T$ and, as standard TEGR, exhibit the violation of  local Lorentz invariance  \cite{FFNTF,  FETMG}.  Extended  teleparallel gravity $f(T)$ differs from  extended gravity $f(R)$ essentially because the former leads to second-order field equations while the latter leads to fourth-order field equations in metric formalism. Then $f(R)$ gravity is not equivalent to $f(T)$ gravity but, if we introduce the boundary terms $B=2\nabla_{\mu}\left(T^{\mu}\right)$, the equivalence can be restored being  $f(T, B)\to f(R)$ gravity \cite{CCC1}.

Extended teleparallel theories $f\left(T\right)$ can be adopted, for example, to explain the accelerated expansion of the Universe at the present time  \cite{Dent}. If we want to study higher-order teleparallel theories, we cannot limit  to $f(T)$, because, as said above, it always produces second-order dynamical equations. To compare extended teleparallel theories with the analogous  in curvature invariants, we have to introduce the boundary term $B$ with its derivatives  and  terms like $\Box T$, $\Box^{k}T$ in the teleparallel Lagrangian \cite{OS, BC}. In order to develop our considerations, we can therefore start from  Lagrangians in terms of $R$ and  $\Box R$ and find their  teleparallel equivalent, using the boundary term $B$. Otherwise we can generally start directly from a theory of gravity of order $n$ in  teleparallel geometry, without the boundary term $B$ but only with $\Box^{k}T$.   We can study, in particular, the field equations associated with the Lagrangian $L_{\Box^{k}T}=e\left(T+a_{0}T^{2}+\sum_{k=1}^{p} a_{k}T\Box^{k}T\right)$.

All the mentioned theories of gravitaty, both in  Riemannian and  teleparallel frameworks,  can be studied in the weak field limit   to be linearized. This allows us to derive gravitational waves (GWs) for each theory  and to analyze their properties. These theories differ in their predictions for propagation speed and polarizations of gravitational radiation. GW polarization measurements, implemented by interferometers,  are a powerful tool to select  viable   theories of gravity. On the other hand, GW detection can contribute to define the effective number of degrees of freedom (d.of.) of a given theory of gravity \cite{CCC1, CCC, CLSX, NPS, Hohmann1, Hohmann2, Hohmann3, LGHL, IKEDA, CCDLV, KPJ, bamba}.  

In this paper, we want to investigate higher-order teleparallel gravity in view of a classification of GW solutions. Furthermore, we want to compare such theories with the analogous in curvature invariants to establish, eventually, if they are equivalent or not.

The layout of the paper is the following.  In Sec.2,   we  obtain the extended teleparallel Lagrangian $L_{\Box T}^{e}$ equivalent to the curvature Lagrangian $L_{\Box R}=-R+a_{0}R^{2}+a_{1}R\Box R$  by means of a boundary term $B$.  As discussed in \cite{Gott,Battaglia,Gott2}, this theory is important because it is related to  one-loop corrections of Quantum Gravity and it gives a double-inflationary evolution in early cosmology because it can be reduced to a double-scalar field theory by conformal transformations.

In Sec.3,  we  obtain the field equations of sixth-order  teleparallel gravity.  Hence, in Sec.4, linearized  equations are derived in the weak field limit.  In Sec.5, adopting the method of  distributions, we get GW solutions  in sixth-order  teleparallel gravity. Properties of GWs,   as  polarization and  helicity,  are investigated by  geodesic deviation and  Newmann-Penrose formalism  in Sec.6.  Sec.7 is devoted to the   study  of  teleparallel gravity at  order $(2p + 2)$. We point out  the non-equivalence  with  extended curvature $R$-based theory considering  a flat geometry with torsion in a zero spin connection. Varying the action $L_{\Box^{k}T}$  with respect to the tetrad $e^{a}_{\phantom{a}\rho}$ and imposing that it vanishes with its derivatives on the boundary of the $\Omega$ domain,  we obtain the teleparallel higher-order field equations  in  presence of matter. After linearization, the related GW solutions are obtained and studied. Conclusions are drawn in Sec.8. In Appendices,    the differential identities and the  variations are reported.

\section{ Teleparallel Equivalent of Sixth Order Curvature Gravity}\label{SOTG}

Let us consider   sixth-order curvature gravity in view of obtaining the teleparallel equivalent  of  Lagrangian $L_{\Box R}=-R+a_{0}R^{2}+a_{1}R\Box R$, expressed in terms of the  Levi Civita connection $\stackrel{\circ}\Gamma$, replacing it with the curvature-free and  torsion-based  Weitzenb\"ock  connection $\tilde{\Gamma}$. In  teleparallelism, we adopt the sixteen components of the tetrad basis $\left\{e_{a}\right\}$ and its dual basis $\left\{e^{a}\right\}$ to describe spacetime manifold. These dynamical variables form a local orthonormal basis for the tangent space at each point  $\{x_{\mu}\}$ of the spacetime. The components of the {\it vierbein} satisfy the relations \cite{LL, CSG, WALD, APIGP, STRAGR, NAK}
\begin{equation}
e^{a}_{\phantom{a}\mu}e_{a}^{\phantom{a}\nu}=\delta_{\mu}^{\nu}\ , \qquad \text{and} \qquad e^{a}_{\ \mu}e_{b}^{\ \mu}=\delta_{a}^{b}\ ,
\end{equation}
\begin{equation}
\eta_{ab}=g_{\mu\nu}e_{a}^{\ \mu}e_{b}^{\phantom{b}\nu}\ ,\qquad \text{and}\qquad g_{\mu\nu}=\eta_{ab}e^{a}_{\phantom{a}\mu}e^{b}_{\phantom{b}\nu}\ ,
\end{equation}
where the signature of the metric $g_{\mu\nu}$ is $\left(+,-,-,-\right)$.
The Weitzenb\"ock connection is defined to  parallel-transport the tetrads on the  manifold, that is
\begin{equation}
\tilde{\Gamma}^{\rho}_{\ \mu\nu}=e_{a}^{\ \rho}\partial_{\nu}e^{a}_{\ \mu}\ ,
\end{equation}
and the related torsion tensor is
\begin{equation}
T^{\nu}_{\phantom{\nu}\rho\mu}=\tilde{\Gamma}^{\nu}_{\phantom{\nu}\mu\rho}-	\tilde{\Gamma}^{\nu}_{\phantom{\nu}\rho\mu}\,.
\end{equation}
Subtracting the Levi-Civita connection from the   Weitzenb\"ock connection, we obtain  the contortion tensor
\begin{equation}
K^{\nu}_{\phantom{\nu} \rho\mu}=\tilde{\Gamma}^{\nu}_{\phantom{\nu} \rho\mu}-\stackrel{\circ}{\Gamma}{\hspace{-0.04in}}^{\nu}_{\phantom{\nu} \rho\mu}=\frac{1}{2}\left(T_{\rho\phantom{\nu} \mu}^{\phantom{\rho}\nu}+T_{\mu\phantom{\mu}\rho}^{\phantom{\mu}\nu}-T^{\nu}_{\phantom{\nu}\rho\mu}\right)\,.
\end{equation}
The superpotential torsion tensor $S^{\rho\mu\nu}$ is
\begin{equation}
S^{\rho\mu\nu}=\frac{1}{2}\left(K^{\mu\nu\rho}-g^{\rho\nu}T^{\sigma\mu}_{\sigma}+g^{\rho\mu}T^{\sigma\nu}_{\sigma}\right)\ .
\end{equation}
Contracting  the torsion tensor with the superpotential,  we get the scalar torsion $T$  
\begin{equation}
T=T_{\rho\mu\nu}S^{\rho\mu\nu}\ .
\end{equation}
On the other hand, the torsion vector  is
\begin{equation}
T^{\sigma}=T^{\nu\sigma}_{\phantom{\nu\sigma}\nu}\ .
\end{equation}  
The  Weitzenb\"ock connection curvature vanishes $R[\tilde{\Gamma}]=0$ but we can express the Levi Civita connection curvature   $R[\stackrel{\circ}\Gamma]$  in terms of the scalar torsion $T$ and the vector torsion $T^{\sigma}$, that is 
\begin{equation}
-R=T+\frac{2}{e}\partial_{\sigma}\left(eT^{\nu\sigma}_{\phantom{\nu\sigma}\nu}\right)\ ,
\end{equation}
and, in more compact form,\footnote{For the signature of the boundary term, see the discussion in \cite{CCT2}.}
\begin{equation}\label{equivalence}
-R=T+B\ ,
\end{equation} 
where $e=det\left(e^{a}_{\phantom{a}\rho}\right)$ and  the boundary term is
\begin{equation}
B=\frac{2}{e}\partial_{\sigma}\left(eT^{\nu\sigma}_{\phantom{\nu\sigma}\nu}\right)\ .
\end{equation}
With the above positions, we can express the  sixth-order curvature Lagrangian in the teleparallel equivalent form by the transformation
\begin{multline}\label{sotle}
-R+a_{0}R^{2}+a_{1}R\Box R= \underbrace{T+a_{0}T^{2}+a_{1}T\Box T}_{\textit{First term}} \\
+\underbrace{a_{0}\left(\frac{2}{e}\partial_{\sigma}\left(eT^{\nu\sigma}_{\phantom{\nu\sigma} \nu}\right)\right)\left(2T+\frac{2}{e}\partial_{\sigma}\left(eT^{\nu\sigma}_{\phantom{\nu\sigma}\nu}\right)\right)}_{\textit{Second Term}}-\underbrace{a_{1}\nabla^{\mu}\left(\frac{2}{e}\partial_{\sigma}\left(eT^{\nu\sigma}_{\phantom{\nu\sigma}\nu}\right)\right)\nabla_{\mu}\left(2T+\frac{2}{e}\partial_{\sigma}\left(eT^{\nu\sigma}_{\phantom{\nu\sigma} \nu}\right)\right)}_{\textit{Third term}}\\
+\underbrace{\frac{1}{e}\partial_{\sigma}\left\{a_{1}h\nabla^{\sigma}\left[\frac{2}{e}\partial_{\eta}\left(eT^{\nu\eta}_{\phantom{\nu\eta} \nu}\right)T+\frac{1}{2}\left(\frac{2}{e}\partial_{\eta}\left(eT^{\nu\eta}_{\phantom{\nu\eta}\nu}\right)\right)^{2}\right]+2eT^{\nu\sigma}_{\phantom{\nu\sigma}\nu}\right\}}_{\textit{Fourth term}}\ ,
\end{multline}
or, more explicitly, via the boundary term $B$
\begin{multline}\label{sotle1}
-R+a_{0}R^{2}+a_{1}R\Box R=T+a_{0}T^{2}+a_{1}T\Box T+a_{0}B\left(2T+B\right)-a_{1}\nabla^{\mu}B\nabla_{\mu}\left(2T+B\right)\\
+\frac{1}{e}\partial_{\sigma}\left\{a_{1}e\nabla^{\sigma}\left[BT+\frac{1}{2}	B^{2}\right]\right\}\ ,
\end{multline}
where $\nabla_{\mu}$ is the covariant derivative with respect to the Levi-Civita connection $\stackrel{\circ}{\Gamma}$.
The fourth term of Eq. \eqref{sotle} is a four-divergence which vanishes if the tetrad with its derivatives vanish on the boundary of the domain; the first term of Eq. \eqref{sotle} is the component we need and the second term   exists because it depends on the $\partial^{2}e_{a}^{\rho}$ of the tetrad and guarantees fourth-order equations in the absence of  $\Box$ operator. Otherwise, considering only  the scalar $T$ depending on the first derivatives of the tetrad $\partial e_{a}^{\rho}$, we cannot go above  the second order (because  $f\left(T\right)$ theories are described  by second order equations and not by fourth-order equations as $f\left(R\right)$ which, instead, depend on the second derivatives of the metric). The third term of Eq. \eqref{sotle}  depends on  third derivatives of  tetrad $\partial^{3}e_{a}^{\rho}$ insuring sixth-order  field equations. The possibility of finding an equivalent teleparallel Lagrangian is guaranteed by the boundary term $B$ which is not incorporated into a four-divergence. The TEGR, whose  Lagrangian depends linearly on the scalar torsion $T$, is equivalent to the Hilbert-Einstein Lagrangian because the term $B$ is incorporated into the four-divergence that vanishes on the boundary while the teleparallel extended theories, to be equivalent to  extended theories like $f\left(R\right)$ or $f\left(R, \Box R\right)$, must include boundary terms. Another form of  sixth-order teleparallel Lagrangian, up to four-divergence, is
\begin{equation}
\mathcal{L}_{\Box T}^{eq}=\underbrace{T+a_{0}B^{2}+a_{1}B\Box B}_{\textit{terms that generate the first order}}+a_{0}T^{2}+a_{1}T\Box T+2a_{0}BT-2a_{1}\nabla^{\mu}B\nabla_{\mu}T-\frac{1}{2}a_{1}\Box B^{2}\ ,
\end{equation}
where we pointed out   parts containing  first-order terms  generating GWs as we are going to discuss in the next sections.

\section{Field Equations of Sixth-Order Teleparallel Equivalent   Gravity}
Let us take into account now   the sixth-order teleparallel equivalent Lagrangian up to a four-divergence \eqref{sotle1}, which can vanish imposing appropriate conditions for the fields on the boundary, that is 
\begin{equation}
L_{\Box T}^{eq}=e\mathcal{L}_{\Box T}^{eq}=\frac{e}{2\kappa^{2}}\left[T+a_{0}T^{2}+a_{1}T\Box T+a_{0}B\left(2T+B\right)-a_{1}\nabla^{\mu}B\nabla{\mu}\left(2T+B\right)\right]\ ,
\end{equation}
where $\kappa^{2}=8\pi G/c^4$ is the gravitational coupling. The following action with the material term $L_{m}=e\mathcal{L}_{m}$ 
\begin{equation}
I=\int_{\Omega}d^{4}x \left[\mathcal{L}_{\Box T}+\mathcal{L}_{m}\right]e\,,
\end{equation}
can be examined.
Varying  with respect to the tetrad $ e^{a}_{ \ \rho}$  with  the variation of the tetrad field vanishing on the boundary, we obtain the following sixth-order field equations:
\begin{equation}
\boxed{
\begin{split}\label{FESOTG}
\frac{4}{e}\partial_{\sigma}\left[e\left(1+2a_{0}T+2a_{1}\Box T\right)S_{a}^{\phantom{a}\rho\sigma}\right]-4\left(1+2a_{0}T+2a_{1}\Box T\right)T^{\mu}_{\phantom{\mu}\nu a}S_{\mu}^{\phantom{\mu}\nu\rho}+\left(T+a_{0}T^{2}+a_{1}T\Box T\right)e_{a}^{\phantom{a}\rho}\\
-a_{1}\biggl\{\left[e_{a}^{\phantom{a}\nu}T\left(\nabla^{\rho}\nabla_{\nu}+\nabla_{\nu}\nabla^{\rho}\right)T-\left(e_{a}^{\phantom{a}\eta}e_{b}^{\phantom{b} \rho}\partial^{\nu}e^{b}_{\phantom{b}\nu}+T_{a}^{\phantom{a}\eta\rho}-e_{a}^{\phantom{a}\eta}T^{\rho}-g^{\eta\rho}T_{a}\right)T\partial_{\eta}T\right]\\
+\frac{1}{e}\partial_{\sigma}\left[e\left(e_{a}^{\phantom{a}\rho}g^{\eta\sigma}-e_{a}^{\phantom{a}\sigma}g^{\eta\rho}-e_{a}^{\phantom{a}\eta}g^{\rho\sigma}\right)T\partial_{\eta}T\right]\biggr\}-\left[a_{0}B^2-a_{1}B\Box\left(T+B\right)+a_{1}\partial^{\mu}B\partial_{\mu}\left(2T+B\right)\right]e_{a}^{\ \rho}\\
+\frac{4}{e}\partial_{\lambda}\left\{e\left[a_{0}\left(T+B\right)+a_{1}\Box \left(T+B\right)\right]\right\}\left(T^{\rho\lambda}_{\phantom{\rho\lambda}a}+e_{a}^{\phantom{a}\lambda}T^{\rho}+g^{\lambda\rho}T_{a}-T^{\lambda}e_{a}^{\phantom{a}\rho}\right)\\
+\frac{4}{e}\partial_{\sigma}\left\{e\partial_{\lambda}\left[a_{0}\left(T+B\right)+a_{1}\Box \left(T+B\right)\right]\left(e_{a}^{\phantom{a}\rho}g^{\lambda\sigma}-e_{a}^{\phantom{a}\sigma}g^{\lambda\rho}\right)\right\}\\
-8\left[a_{0}B+a_{1}\Box B\right]\left(T^{\mu}_{\phantom{\mu}\nu a}S_{\mu}^{\phantom{\mu} \nu\rho}\right)+\frac{8}{e}\partial_{\sigma}\left[e\left(a_{0}B+a_{1}\Box B\right)S_{a}^{\phantom{a}\rho\sigma}\right]=2\kappa^{2}\mathcal{T}^{\left(m\right)\phantom{a}\rho}_{\phantom{\left(m\right)}a}
\end{split}
}\ ,
\end{equation}
with $T^{\lambda}=T^{\nu\lambda}_{\phantom{\nu\lambda} \nu}$ or alternatively:
\begin{multline}
\frac{4}{e}\partial_{\sigma}\left[e\left(1+2a_{0}T+2a_{1}\Box T\right)S_{a}^{\ \rho\sigma}\right]-4\left(1+2a_{0}T+2a_{1}\Box T\right)T^{\mu}_{\phantom{\mu}\nu a}S_{\mu}^{\phantom{\mu}\nu\rho}+\left(T+a_{0}T^{2}+a_{1}T\Box T\right)e_{a}^{\phantom{a}\rho}\\
-a_{1}\Biggl\{\left[e_{a}^{\phantom{a}\nu}T\left(\nabla^{\rho}\nabla_{\nu}+\nabla_{\nu}\nabla^{\rho}\right)T-\left(e_{a}^{\phantom{a}\eta}e_{b}^{\phantom{b}\rho}\partial^{\nu}e^{b}_{\phantom{b}\nu}+T_{a}^{\phantom{a}\eta\rho}-e_{a}^{\phantom{a}\eta}T^{\rho}-g^{\eta\rho}T_{a}\right)T\partial_{\eta}T\right]\\
+\frac{1}{e}\partial_{\sigma}\left[e\left(e_{a}^{\phantom{a}\rho}g^{\eta\sigma}-e_{a}^{\phantom{a}\sigma}g^{\eta\rho}-e_{a}^{\phantom{a}\eta}g^{\rho\sigma}\right)T\partial_{\eta}T\right]\Biggr\}-\left[a_{0}B^2-a_{1}B\Box\left(T+B\right)+a_{1}\partial^{\mu}B\partial_{\mu}\left(2T+B\right)\right]e_{a}^{\phantom{a}\rho}
\\
+\frac{4}{e}\partial_{\lambda}\left(e G \right)\left(T^{\rho\lambda}_{\phantom{\rho\lambda}a}+e_{a}^{\phantom{a}\lambda}T^{\rho}+g^{\lambda\rho}T_{a}-T^{\lambda}e_{a}^{\phantom{a}\rho}\right)
+\frac{4}{e}\partial_{\sigma}\left[e\partial_{\lambda}G\left(e_{a}^{\phantom{a}\rho}g^{\lambda\sigma}-e_{a}^{\phantom{a}\sigma}g^{\lambda\rho}\right)\right]
\\
-8\left[a_{0}B+a_{1}\Box B\right]\left(T^{\mu}_{\ \nu a}S_{\mu}^{\phantom{\mu}\nu\rho}\right)+\frac{8}{e}\partial_{\sigma}\left[e\left(a_{0}B+a_{1}\Box B\right)S_{a}^{\phantom{a}\rho\sigma}\right]=2\kappa^{2}\mathcal{T}^{\left(m\right)\phantom{a}\rho}_{\phantom{\left(m\right)}a}\ ,
\end{multline}
where $G=a_{0}\left(T+B\right)+a_{1}\Box\left(T+B\right)$. The energy-momentum tensor of matter $\mathcal{T} ^{\left(m\right)\phantom{a}\rho}_{\phantom{\left(m\right)}a}$ is:
\begin{equation}
\mathcal{T}^{\left(m\right)\phantom{a}\rho}_{\phantom{\left(m\right)}a}=-\frac{1}{e}\frac{\delta\left(e\mathcal{L}_{m}\right)}{\delta e^{a}_{\phantom{a}\rho}}\ .
\end{equation}
For  details on  differential identities  and variations, see Appendices  \ref{A} and \ref{B}.

\section{Weak Field Limit of  Sixth-Order Teleparallel Equivalent Gravity}
Let us  perturb the tetrad field around the background metric  with the hypothesis that we are  sufficiently far from  massive bodies.  We can limit our considerations to a flat background  described by the trivial tetrad $e^{a}_{\phantom{a}\mu}=\delta^{a}_{\phantom{a}\mu}$ as follows
\begin{equation}
e^{a}_{\phantom{a}\mu}=\delta^{a}_{\phantom{a}\mu}+E^{a}_{\phantom{a}\mu}\ ,
\end{equation}
where $\lvert E^{a}_{\phantom{a}\mu} \rvert \ll 1$. 
We therefore expand the metric tensor $g_{\mu\nu}$  to first order  in $E^{a}_{\phantom{a}\mu}$ and we get 
\begin{equation}
g_{\mu\nu}=\eta_{\mu\nu}+h_{\mu\nu}+\mathcal{O}\left(h^2\right)=\eta_{\mu\nu}+\eta_{\mu a}E^{a}_{\phantom{a}\nu}+\eta_{\nu a}E^{a}_{\phantom{a}\mu}+\mathcal{O}\left(E^2\right)\,.
\end{equation}
Perturbations are related each other by the relation
\begin{equation}
h_{\mu\nu}=\eta_{\mu a}E^{a}_{\phantom{a}\nu}+\eta_{\nu a}E^{a}_{\phantom{a}\mu}\ .
\end{equation}
To first order in $E^{a}_{\phantom{a}\mu}$, the Weitzenb\"ock connection becomes
\begin{equation}
\tilde{\Gamma}^{\rho\left(1\right)}_{\phantom{\rho}\mu\nu}=\delta_{a}^{\phantom{a}\rho}\partial_{\nu}E^{a}_{\phantom{a}\mu}\,.
\end{equation}
The zero-order covariant derivative $\nabla_{\mu}$ and the covariant d'Alembert operator $\Box=g^{\mu\nu}\nabla_{\mu}\nabla_{\nu}$ become
\begin{equation}
\nabla_{\mu}^{\left(0\right)}=\partial_{\mu}\ ,
\end{equation}
\begin{equation}
\Box^{\left(0\right)}=\eta^{\mu\nu}\partial_{\nu}\partial_{\mu}=\partial^{\mu}\partial_{\mu}\ .
\end{equation}
The torsion tensor $T^{\mu}_{\phantom{\mu}\nu\rho}$, its contraction $T^{\mu\nu}_{\phantom{\mu\nu}\mu}$ and the contortion tensor $K^{\rho}_{\mu\nu}$, to first order in tetrad perturbation $E^{a}_{\phantom{a}\mu}$ can be written as 
\begin{equation}
T^{\mu\left(1\right)}_{\phantom{\mu}\nu\rho}=\delta_{a}^{\phantom{a}\mu}\left(\partial_{\nu}E^{a}_{\phantom{a}\rho}-\partial_{\rho}E^{a}_{\phantom{a}\nu}\right)\ ,
\end{equation}
\begin{equation}
T^{\rho\sigma\left(1\right)}_{\phantom{\rho\sigma}\rho}=\delta_{a}^{\phantom{a}\mu}\eta^{\sigma\nu}\left(\partial_{\nu}E^{a}_{\phantom{a}\mu}-\partial_{\mu}E^{a}_{\phantom{a}\nu}\right)\ ,
\end{equation}
\begin{equation}
K^{\rho\left(1\right)}_{\mu\nu}=\eta_{\mu\lambda}\delta^{\phantom{a}\lambda}_{a}\partial^{\rho}E^{a}_{\phantom{a}\nu}-\delta_{a}^{\phantom{a}\rho}\partial_{\mu}E^{a}_{\phantom{a}\nu}\ .
\end{equation} 
The superpotential $S_{\rho}^{\phantom{\rho}\mu\nu}$, the scalar torsion $T$ and the boundary term $B$, perturbed at the lowest order, become 
\begin{align}
2S_{\rho}^{\phantom{\rho}\mu\nu\left(1\right)}=& \delta_{a}^{\phantom{a}\nu}\partial^{\mu}E^{a}_{\phantom{a}\rho}-\delta_{a}^{\phantom{a}\mu}\partial^{\nu}E^{a}_{\phantom{a}\rho}-\delta^{\nu}_{\rho}\left(\delta_{a}^{\phantom{a}\sigma}\partial^{\mu}E^{a}_{\phantom{a}\sigma}-\eta^{\alpha\mu}\delta^{a}_{\phantom{a}\alpha}\partial_{\sigma}E^{\sigma}_{\phantom{\sigma}a}\right)\nonumber\\
&+\delta^{\mu}_{\rho}\left(\delta_{a}^{\phantom{a}\sigma}\partial^{\nu}E^{a}_{\phantom{a}\sigma}-\eta^{\alpha\nu}\delta^{a}_{\phantom{a}\alpha}\partial_{\sigma}E^{\sigma}_{\phantom{\sigma}a}\right)\ ,
\end{align}
\begin{equation}
T^{\left(2\right)}=T^{\mu\nu\rho\left(1\right)}S^{\left(1\right)}_{\mu\nu\rho}\ ,
\end{equation}
\begin{equation}
B^{\left(1\right)}=\left(\frac{2}{e}\partial_{\sigma}\left(e T^{\sigma}\right)\right)^{\left(1\right)}=2\delta_{a}^{\phantom{a}\nu}\left[\Box E^{a}_{\phantom{a}\nu}-\partial^{\mu}\partial_{\nu}E^{a}_{\phantom{a}\mu}\right]\,.
\end{equation}
The lowest order where the scalar torsion $T$ does not vanish is the second order in the tetrad perturbation $E^{a}_{\phantom{a}\mu}$. The Ricci curvature $R$, to first order in $E^{a}_{\phantom{a}\mu}$, gets the following form
\begin{equation}
R^{\left(1\right)}=-B^{\left(1\right)}\ ,
\end{equation}
that is, to first order, only the boundary term $B$ contributes to the Ricci curvature and not the torsion $T$. Finally we obtain the useful relation 
\begin{align}\label{DerSupPot}
2\partial_{\nu}S_{\rho}^{\phantom{\rho}\mu\nu(1)}=&\delta_{a}^{\phantom{a}\nu}\partial_{\nu}\partial^{\mu}E^{a}_{\phantom{a}\rho}-\delta_{a}^{\phantom{a}\mu}\Box E^{a}_{\phantom{a}\rho}-\delta_{a}^{\phantom{a}\sigma}\partial_{\rho}\partial^{\mu}E^{a}_{\phantom{a}\sigma} \nonumber\\
&+\eta^{\rho\mu}\delta_{a}^{\phantom{a}\sigma}\partial_{\rho}\partial_{\sigma}E^{a}_{\phantom{a}\rho}+\delta^{\mu}_{\rho}\delta_{a}^{\phantom{a}\sigma}\Box E^{a}_{\phantom{a}\sigma}-\delta^{\mu}_{\rho}\delta_{a}^{\phantom{a}\sigma}\eta^{\alpha\nu}\partial_{\nu}\partial_{\sigma}E^{a}_{\phantom{a}\alpha}\ .
\end{align}
The first-order tetrad perturbation $E^{a}_{\phantom{a}\mu}$ is not symmetric because the sixth-order teleparallel equivalent gravity is not invariant under a local Lorentz transformation \cite{AC, OP}
\begin{equation}
\eta_{\mu a}E^{a}_{\phantom{a}\nu}\neq \eta_{\nu a}E^{a}_{\phantom{a}\mu}\ ,
\end{equation}
and then, we can decompose the tetrad perturbation $E_{\mu\nu}$ into symmetric and antisymmetric parts
\begin{equation}
 E_{\mu\nu}=E_{\left(\mu\nu\right)}+E_{\left[\mu\nu\right]}\ .
\end{equation}  
However, the  antisymmetric term  $E_{\left[\mu\nu\right]}$   has no physical meaning because it is not involved 
into the Lagrangian $L_{\Box T}^{eq}$ and the  field equations: in fact,  they depend on the symmetric term $E_{\left(\mu\nu\right)}$ by means of $T$ and $B$.
Hence, we can set to zero the antisymmetric part $E_{\left[\mu\nu\right]}$
\begin{equation}
E_{\left[\mu\nu\right]}=0\ ,
\end{equation}
and the metric perturbation becomes 
\begin{equation}
h_{\mu\nu}=2\eta_{\mu a}E^{a}_{\phantom{a}\nu}\ .
\end{equation}
We now linearize the field equations of sixth-order teleparallel equivalent gravity, Eq.\eqref{FESOTG}, keeping, at  most,  first order terms in tetrad perturbation $E^{a}_{\phantom{a}\mu}$. We therefore obtain
\begin{equation}\label{WFLEFSO}
4\partial_{\sigma}S_{\tau}^{\phantom{\tau}\rho\sigma\left(1\right)}+4\left(\eta^{\lambda\sigma}\delta^{\rho}_{\tau}-\eta^{\lambda\rho}\delta^{\sigma}_{\tau}\right)\partial_{\sigma}\partial_{\lambda}\left(a_{0}B^{\left(1\right)}+a_{1}\Box B^{\left(1\right)}\right)=2\kappa^{2}\mathcal{T}_{\tau}^{\phantom{\tau}\rho\left(0\right)}\ .
\end{equation}
The harmonic gauge, to the  first order in tetrad perturbation, becomes
\begin{equation}
\partial_{\mu}\bar{E}^{\mu\nu}=\partial_{\mu}\left(E^{\mu\nu}-\frac{1}{2}\eta^{\mu\nu}E\right)=0\ ,
\end{equation}
where we  set
\begin{equation}
E_{\mu\nu}=\eta_{\mu a}E^{a}_{\phantom{a}\nu}\ , \qquad E=\delta_{a}^{\phantom{a}\mu}E^{a}_{\phantom{a}\mu}\ ,
\end{equation}
and
\begin{equation}\label{newfieldgaugelorentz}
\bar{E}_{\mu\nu}=E_{\mu\nu}-\frac{1}{2}\eta_{\mu\nu}E\ .
\end{equation}
The boundary term to  first order $B^{\left(1\right)}$ and  Eq.\eqref{DerSupPot} in the harmonic gauge take the form
\begin{equation}\label{B1Or}
B^{\left(1\right)}=-\Box\bar{E}\ ,
\end{equation}
\begin{equation}\label{2DS1O}
2\partial_{\nu}S_{\rho}^{\phantom{\rho}\mu\nu(1)}=-\Box\bar{E}^{\mu}_{\phantom{\rho}\rho}\ .
\end{equation}
Substituting  Eqs.\eqref{B1Or} and \eqref{2DS1O} into \eqref{WFLEFSO}, we obtain the linearized field equations in presence of matter \eqref{WFLEFSO} as
\begin{equation}
\boxed{
\Box\bar{E}^{\rho}_{\phantom{\rho}\tau}+2a_{0}\left(\delta^{\rho}_{\tau}\Box^{2}-\partial_{\tau}\partial^{\rho}\Box\right)\bar{E}+2a_{1}\left(\delta^{\rho}_{\tau}\Box^{3}-\partial_{\tau}\partial^{\rho}\Box^{2}\right)\bar{E}=-\kappa^{2}\mathcal{T}^{\phantom{\tau}\rho\left(0\right)}_{\tau}
}\ .
\end{equation}
In a more compact form 
\begin{equation}\label{LinearEqfield}
\boxed{
\Box\bar{E}^{\rho}_{\phantom{\rho}\tau}+2\sum_{k=0}^{1}a_{k}\left(\delta^{\rho}_{\tau}\Box^{k+2}-\partial_{\tau}\partial^{\rho}\Box^{k+1}\right)\bar{E}=-\kappa^{2}\mathcal{T}^{\phantom{\tau}\rho\left(0\right)}_{\tau}
}\ ,
\end{equation}
where the trace is
\begin{equation}
\Box\bar{E}+6\sum_{k=0}^{1}a_{k}\Box^{k+2}\bar{E}=-\kappa^{2}\mathcal{T}^{\left(0\right)}\ .
\end{equation}
Setting $l=k+2$ and defining $c_{l}$ as
\begin{equation*}
  c_{l}:=
\begin{cases}
     1 & \text{if $l=1$}\\
     6a_{l-2} & \text{if $l>1$}
\end{cases}\ ,
\end{equation*}
we get the trace equation of Eq. \eqref{LinearEqfield} as
\begin{equation}\label{Traceeq}
\boxed{
\sum_{l=1}^{3}c_{l}\Box^{l}\bar{E}=-\kappa^{2}\mathcal{T}^{\left(0\right)}
}\ .
\end{equation}
We have now all the ingredient to search for GW solutions in sixth-order teleparallel gravity.

\section{Gravitational Waves in Sixth-Order Teleparallel Equivalent Gravity}
The solution of  Eq.\eqref{Traceeq} is an object more general than a function. $\bar{E}\left(x\right)$ is a distribution in $\mathbb{R}^{4}$ \cite{GERVANDIJK} or more precisely a {\it tempered distribution}. 

Let us now give a precise definition of the notion of  distribution. Let $\phi$ be a complex-valued function on $\mathbb{R}^{4}$. The closure of the set of points $\left\{x\in\mathbb{R}^{4}: \phi(x) \neq 0\right\}$ is called the support of $\phi$ and it is denoted by $supp(\phi)$. A function $\phi: \mathbb{R}^{4}:\to\mathbb{C}$ is called a $C^{\infty}$ function if all its partial derivatives $D^{k}\phi$ exist and are continuous. A $C^{\infty}$ function with compact support is called a {\it test function} and the space of the test functions on $\mathbb{R}^{4}$ is denoted by $\mathcal{D}\left(\mathbb{R}^{4}\right)$ that is, the space of functions with compact support. A distribution on $\mathbb{R}^{4}$ is a continuous complex-valued linear functional $T$ defined on $\mathcal{D}\left(\mathbb{R}^{4}\right)$ and the space of all distributions $T$ is denoted by $\mathcal{D}'\left(\mathbb{R}^{4}\right)$. The space $\mathcal{D}\left(\mathbb{R}^{4}\right)$ is however not closed under Fourier transforms thus we replace it with a new space, closed under Fourier transforms, the Schwartz  space $\mathcal{S}(\mathbb{R}^{4})$ of rapidly decreasing functions. The space of Schwartz functions is defined as the space of infinitely differentiable functions whose derivatives decay faster than any polynomial at infinity, that is, formally, $\mathcal{S}(\mathbb{R}^{4})=\left\{\phi\in C^{\infty}:\Vert\phi\Vert_{\left(N,\alpha\right)}<\infty\  \forall N,\alpha\right\}$ where we have defined the norm $\Vert\phi\Vert_{\left(N,\alpha\right)}=\sup_{x\in\mathbb{R}^{4}}\left(1+\vert x\vert^{N}\right)\vert\partial^{\alpha}\phi\left(x\right)\vert$ for any non-negative integer $N$ and any multi-index $\alpha$. 
The tempered distribution is a continuous linear functional on $\mathcal{S}(\mathbb{R}^{4})$ and the space of all tempered distributions on
$\mathbb{R}^{4}$ is denoted by $\mathcal{S}'(\mathbb{R}^{4})$, that is the dual space of the space of Schwartz functions $\mathcal{S}(\mathbb{R}^{4})$. 

The equation of  trace  \eqref{Traceeq} in vacuum, in $k$-space, becomes
\begin{equation}\label{Tracecekspace}
\left(\sum_{l=1}^{3}c_{l}\left(-1\right)^{l}k^{2l}\right)\hat{A}\left(k\right)=0\ ,
\end{equation}
where $k^{2}=\omega^{2}-\mathbf{k}\cdot\mathbf{k}=\omega^{2}-q^{2}$ with the four-wavevector $k^{\mu}=\left(\omega,\bf{k}\right)$.  $\hat{A}\left(k\right)\in\mathcal{S}'(\mathbb{R}^{4})$ is the Fourier transformation of $\bar{E}\left(x\right)\in\mathcal{S}'(\mathbb{R}^{4})$
\begin{equation}
\hat{A}\left(k\right)=\int\frac{d^{4}k}{\left(2\pi\right)^{2}}\bar{E}\left(x\right)e^{-ik^{\alpha}x_{\alpha}}\ .
\end{equation}
 We solve Eq.\eqref{Tracecekspace} as an algebraic equation in the space of distributions with $\hat{A}\left(k\right)\in\mathcal{S}'(\mathbb{R}^{4})$ 
\begin{equation}
\left(6a_{1}k^{6}-6a_{0}k^{4}+k^{2}\right)\hat{A}\left(k\right)=0\ ,
\end{equation} 
then 
\begin{equation}
\hat{A}\left(k\right)=f\left(k\right)\left[\delta\left(6a_{1}k^{6}-6a_{0}k^{4}+k^{2}\right)\right]\ ,
\end{equation}
where we used the $\delta$-distribution with a suitable complex function $f(k)$. We use  now the following properties of $\delta$-distribution 
\begin{equation}\label{Prodelta1}
\delta\left(F\left(x\right)\right)=\sum_{h=1}^{n}\frac{\delta\left(x-x_{h}\right)}{\vert F'\left(x_{h}\right)\vert}\ ,
\end{equation}
such as $F(x_{h})=0$ for $h=1,\dots,n$ and 
\begin{equation}\label{Prodelta2}
f\left(x\right)\delta\left(x-a\right)=f\left(a\right)\delta\left(x-a\right)\ .
\end{equation}
The solutions of the related algebraic equation 
\begin{equation}\label{algebraicequation}
6a_{1}k^{6}-6a_{0}k^{4}+k^{2}=0\ ,
\end{equation}
are
\begin{equation}
k^{2}=0\qquad k^{2}=M_{2}^{2}\neq 0\qquad k^{2}=M_{3}^{2}\neq 0\ ,
\end{equation}
where
\begin{equation}
M_{2,3}^{2}=\frac{-3a_{0}\mp\sqrt{3a_{0}^{2}-a_{1}}}{6a_{1}}\ .
\end{equation}
Keeping $\mathbf{k}$ fixed and varying the time-component $k^{0}$ of four-wavevector $k^{\mu}$,  we obtain 
\begin{equation}
k^{2}=0\rightarrow (k^{0})^{2}=\vert\mathbf{k}\vert^{2}\rightarrow k^{0}=\pm\vert\mathbf{k}\vert=\pm\omega_{1}\ ,
\end{equation}
\begin{equation}
k^{2}=M_{2}^{2}\rightarrow\left(k_{0}\right)^{2}-\vert\mathbf{k}\vert^{2}=M_{2}^{2}\rightarrow k^{0}=\pm\sqrt{M_{2}^{2}+\vert\mathbf{k}\vert^{2}}=\pm\omega_{2}\ ,
\end{equation}
\begin{equation}
k^{2}=M_{3}^{2}\rightarrow\left(k_{0}\right)^{2}-\vert\mathbf{k}\vert^{2}=M_{3}^{2}\rightarrow k^{0}=\pm\sqrt{M_{3}^{2}+\vert\mathbf{k}\vert^{2}}=\pm\omega_{3}\ .
\end{equation}
From \eqref{Prodelta1}, it is 
\begin{align}
\delta\left(6a_{1}k^{6}-6a_{0}k^{4}+k^{2}\right)=&\frac{1}{2\omega_{1}\vert 18a_{1}\omega_{1}^{4}-12a_{0}\omega_{1}^{2}+1\vert}\left[\delta\left(k^{0}-\omega_{1}\right)+\delta\left(k^{0}+\omega_{1}\right)\right]\nonumber\\
&+\frac{1}{2\omega_{2}\vert 18a_{1}\omega_{2}^{4}-12a_{0}\omega_{2}^{2}+1\vert}\left[\delta\left(k^{0}-\omega_{2}\right)+\delta\left(k^{0}+\omega_{2}\right)\right]\\
&+\frac{1}{2\omega_{3}\vert 18a_{1}\omega_{3}^{4}-12a_{0}\omega_{3}^{2}+1\vert}\left[\delta\left(k^{0}-\omega_{3}\right)+\delta\left(k^{0}+\omega_{3}\right)\right]\nonumber\ ,
\end{align}
and we have
\begin{equation}
\hat{A}\left(k\right)=f\left(k^{0},\mathbf{k}\right)\sum_{m=1}^{3}\frac{\delta\left(k^{0}-\omega_{m}\right)+\delta\left(k^{0}+\omega_{m}\right)}{2\omega_{m}\vert 18a_{1}\omega_{m}^{4}-12a_{0}\omega_{m}^{2}+1\vert}\ ,
\end{equation}
and, by  \eqref{Prodelta2}, we get
\begin{equation}\label{Amplitudesmomentum}
\hat{A}\left(k\right)=\sum_{m=1}^{3}\frac{\delta\left(k^{0}-\omega_{m}\right)f\left(\omega_{m},\mathbf{k}\right)+\delta\left(k^{0}+\omega_{m}\right)f\left(-\omega_{m},\mathbf{k}\right)}{2\omega_{m}\vert 18a_{1}\omega_{m}^{4}-12a_{0}\omega_{m}^{2}+1\vert}\ .
\end{equation}
Imposing that  the function $\bar{E}\left(x\right)$ is real,  we obtain
\begin{equation}
f^{*}\left(k\right)=f\left(-k\right)\ ,
\end{equation}
that is
\begin{equation}\label{relation1}
f^{*}\left(\omega_{m},\mathbf{k}\right)=f\left(-\omega_{m},-\mathbf{k}\right)\rightarrow f^{*}\left(\omega_{m},-\mathbf{k}\right)=f\left(-\omega_{m},\mathbf{k}\right)\ .
\end{equation}
After setting
\begin{equation}\label{setting1}
f\left(\omega_{m},\mathbf{k}\right)=Q_{m}\left(\mathbf{k}\right)\ ,
\end{equation}
from \eqref{relation1} and \eqref{setting1},   we obtain, from Eq. \eqref{Amplitudesmomentum},
\begin{equation}
\hat{A}\left(k\right)=\sum_{m=1}^{3}\frac{\delta\left(k^{0}-\omega_{m}\right)Q_{m}\left(\mathbf{k}\right)+\delta\left(k^{0}+\omega_{m}\right)Q_{m}^{*}\left(-\mathbf{k}\right)}{2\omega_{m}\vert 18a_{1}\omega_{m}^{4}-12a_{0}\omega_{m}^{2}+1\vert}\ .
\end{equation}
In a  more compact form,  it is 
\begin{equation}\label{AkB}
\hat{A}\left(k\right)=\sqrt{2\pi}\sum_{m=1}^{3}\left[\delta\left(k^{0}-\omega_{m}\right)\hat{B}_{m}\left(\mathbf{k}\right)+\delta\left(k^{0}+\omega_{m}\right)\hat{B}_{m}^{*}\left(-\mathbf{k}\right)\right]\ ,
\end{equation}
with 
\begin{equation}
\hat{B}_{m}\left(\mathbf{k}\right)=\frac{Q_{m}\left(\mathbf{k}\right)}{2\sqrt{2\pi}\omega_{m}\vert 18a_{1}\omega_{m}^{4}-12a_{0}\omega_{m}^{2}+1\vert}\ .
\end{equation}
Let us  now perform the inverse Fourier transform of $\hat{A}\left(k\right)$. It  is
\begin{equation}
\bar{E}\left(x\right)=\int \frac{d^{4}k}{\left(2\pi\right)^{2}}\hat{A}\left(k\right)e^{ik^{\alpha}x_{\alpha}}\ ,
\end{equation}
and, by \eqref{AkB}, we get
\begin{equation}
\bar{E}\left(x\right)=\frac{1}{\left(2\pi\right)^{3/2}}\int d^{3}\mathbf{k}\int dk^{0}e^{i\left(k^{0}x^{0}-\mathbf{k}\cdot\mathbf{x}\right)}\sum_{m=1}^{3}\left[\delta\left(k^{0}-\omega_{m}\right)\hat{B}_{m}\left(\mathbf{k}\right)+\delta\left(k^{0}+\omega_{m}\right)\hat{B}_{m}^{*}\left(-\mathbf{k}\right)\right]\ .
\end{equation}
Finally, we obtain the solution 
\begin{equation}
\bar{E}\left(x\right)=\frac{1}{\left(2\pi\right)^{3/2}}\sum_{m=1}^{3}\int d^{3}\mathbf{k}\left(\hat{B}_{m}\left(\mathbf{k}\right)e^{ik_{m}^{\alpha}x_{\alpha}}+c.c.\right)\ ,
\end{equation}
with $k_{m}^{\mu}=\left(\omega_{m},\mathbf{k}\right)$. Let us rename $M_{2}$ as $k_{2}$ and $M_{3}$ as $k_{3}$, then we have, from \eqref{algebraicequation},
\begin{equation}\label{solutionk123}
\Biggl\{
\begin{array}{ll}
k_{1}^{2}=0  & \mbox{if $m=1$}\\
\sum_{l=0}^{1}a_{l}\left(-1\right)^{l+2}k_{m}^{2\left(l+1\right)}=\frac{1}{6} & \mbox{if $m=2,3$}
\end{array}\ .
\end{equation}
Therefore, from Eqs.\eqref{LinearEqfield} and \eqref{solutionk123},  we get, in vacuum,  
\begin{equation}
\Box\bar{E}_{\rho\tau}\left(x\right)=\sum_{m=2}^{3}\int \frac{d^{3}\mathbf{k}}{(2\pi)^{3/2}}\left\{\left(-\frac{k_{m}^{2}}{3}\right)\left[\eta_{\rho\tau}-\frac{\left(k_{m}\right)_{\rho}\left(k_{m}\right)_{\tau}}{k_{m}^{2}}\right]\right\}\left(\hat{B}_{m}\left(\mathbf{k}\right)e^{ik_{m}^{\alpha}x_{\alpha}}+c.c.\right)\ .
\end{equation}
The general solution of \eqref{LinearEqfield}, in vacuum,  considering the  homogeneous plus a particular solution is
\begin{align}
\bar{E}_{\rho\tau}\left(x\right)&=\int \frac{d^{3}\mathbf{k}}{(2\pi)^{3/2}}\hat{C}_{\rho\tau}\left(\mathbf{k}\right)e^{ik_{1}^{\alpha}x_{\alpha}} \nonumber\\
&+\sum_{m=2}^{3}\int \frac{d^{3}\mathbf{k}}{(2\pi)^{3/2}}\left\{\left(-\frac{1}{3}\right)\left[\eta_{\rho\tau}-\frac{\left(k_{m}\right)_{\rho}\left(k_{m}\right)_{\tau}}{k_{m}^{2}}\right]\right\}\hat{B}_{m}\left(\mathbf{k}\right)e^{ik_{m}^{\alpha}x_{\alpha}}+c.c.\ .
\end{align}
From Eq.\eqref{newfieldgaugelorentz}, we derive  the GWs in vacuum for sixth-order teleparallel equivalent gravity, that is
\begin{align}\label{GWEFTBTG}
E_{\rho\tau}\left(x\right)&=\int \frac{d^{3}\mathbf{k}}{(2\pi)^{3/2}}C_{\rho\tau}\left(\mathbf{k}\right)e^{ik_{1}^{\alpha}x_{\alpha}} \nonumber\\
&+\sum_{m=2}^{3}\int \frac{d^{3}\mathbf{k}}{(2\pi)^{3/2}}\left\{\frac{1}{3}\left[\frac{\eta_{\rho\tau}}{2}+\frac{\left(k_{m}\right)_{\rho}\left(k_{m}\right)_{\tau}}{k_{m}^{2}}\right]\right\}\hat{B}_{m}\left(\mathbf{k}\right)e^{ik_{m}^{\alpha}x_{\alpha}}+c.c.\ .
\end{align}
For $f\left(T, B\right)$ teleparallel gravity and  for higher-order metric gravity see \cite{CCC1, CCC}.

\section{Polarizations and helicity}\label{POLHEL}
Let us  consider the wave traveling along the $+z$ direction in a local proper reference frame with a separation vector $\vec{\chi}=(x^{1}, x^{2}, x^{3})$ which connects two nearby geodesics.  The geodesic deviation is
\begin{equation}\label{eqdevgeoelectric}
\ddot x^{i}=-R^{i}_{\phantom{i}0k0}x^{k}\ ,
\end{equation}
where the Latin index ranges over the set $\left\{1,2,3\right\}$ and $R^{i}_{\phantom{i}0k0}$ are the so-called "electric" components of the Riemann tensor, the only measurable components.
Linearized electric components of the Riemann tensor $R^{\left(1\right)}_{\phantom{1}i0j0}$, expressed in terms of the tetrad perturbation $E_{\mu\nu}$, restricted to the Hilbert space of  square integrable functions denoted by ${L}^{2}\left(\mathbb{R}^{4}\right)$, are 
\begin{equation}
R^{\left(1\right)}_{\phantom{1}i0j0}=\left(E_{i0,0j}+E_{0j,i0}-E_{ij,00}-E_{00,ij}\right)\ .
\end{equation}
Replacing them into Eq.\eqref{eqdevgeoelectric},  we obtain
\begin{equation}\label{eqdevgeolinear}
\begin{cases}
\ddot x(t)=-\left(xE_{11,00}+yE_{12,00}\right) \\
\ddot y(t)=- \left(xE_{12,00}+yE_{11,00}\right)\\
\ddot z(t)=\left(2E_{03,03}-E_{33,00}-E_{00,33}\right)z
\end{cases}\ .
\end{equation}

From the general solution \eqref{GWEFTBTG}, the plane wave propagating in the $+\hat{z}$ direction with $k_{1}^{2}=0$, keeping $\mathbf{k}$ fixed and $k_{1}^{\mu}=\left(\omega_{1},0,0,k_{z}\right)$ can be written as 
\begin{equation}\label{FirstOrderTetradmassless}
E^{(k_{1})}_{\mu\nu}\left(t,z\right)=\sqrt{2}\left[\hat{\epsilon}^{(+)}\left(\omega_{1}\right)\epsilon^{(+)}_{\mu\nu}+\hat{\epsilon}^{(\times)}\left(\omega_{1}\right)\epsilon^{(\times)}_{\mu\nu}\right]e^{i\omega_{1}\left(t-z\right)}+c.c.\ ,
\end{equation}
where 
\begin{equation}
\epsilon^{(+)}_{\mu\nu}=\frac{1}{\sqrt{2}}
\begin{pmatrix} 
0 & 0 & 0 & 0 \\
0 & 1 & 0 & 0 \\
0 & 0 & -1 & 0 \\
0 & 0 & 0 & 0
\end{pmatrix}\ ,
\end{equation}
\begin{equation}
\epsilon^{(\times)}_{\mu\nu}=\frac{1}{\sqrt{2}}
\begin{pmatrix} 
0 & 0 & 0 & 0 \\
0 & 0 & 1 & 0 \\
0 & 1 & 0 & 0 \\
0 & 0 & 0 & 0
\end{pmatrix}\ .
\end{equation}
Always from Eq.\eqref{GWEFTBTG}, the $m$-th plane wave propagating in the $+\hat{z}$ direction, for $k_{m}^{2}\neq 0$ with $m\in\left\{2,3\right\}$, keeping $\mathbf{k}$ fixed,  and $k_{m}^{\mu}=\left(\omega_{m},0,0,k_{z}\right)$ can be written as 
\begin{multline}\label{FirstOrderTetradmassive}
E^{(k_{m})}_{\mu\nu}\left(t,z\right)=\Biggl[\left(\frac{1}{2}+\frac{\omega^{2}_{m}}{k_{m}^{2}}\right)\epsilon_{\mu\nu}^{(TT)}-\frac{\sqrt{2}\omega_{m}k_{z}}{k_{m}^{2}}\epsilon_{\mu\nu}^{(TS)}\\
-\frac{1}{\sqrt{2}}\epsilon_{\mu\nu}^{(b)}+\left(-\frac{1}{2}+\frac{k_{z}^{2}}{k_{m}^{2}}\right)\epsilon_{\mu\nu}^{(l)}\Biggr]\frac{\hat{B}_{m}\left(k_{z}\right)}{3}e^{i\left(\omega_{m}t-k_{z}z\right)}+c.c.\ ,
\end{multline}
where 
\begin{align}
\epsilon^{(TT)}_{\mu\nu}&=
\begin{pmatrix} 
1 & 0 & 0 & 0 \\
0 & 0 & 0 & 0 \\
0 & 0 & 0 & 0 \\
0 & 0 & 0 & 0
\end{pmatrix}\ , &
\epsilon^{(TS)}_{\mu\nu}&=\frac{1}{\sqrt{2}}
\begin{pmatrix} 
0 & 0 & 0 & 1 \\
0 & 0 & 0 & 0 \\
0 & 0 & 0 & 0 \\
1 & 0 & 0 & 0
\end{pmatrix}\ ,\\
\epsilon^{(b)}_{\mu\nu}&=\frac{1}{\sqrt{2}}
\begin{pmatrix} 
0 & 0 & 0 & 0 \\
0 & 1 & 0 & 0 \\
0 & 0 & 1 & 0 \\
0 & 0 & 0 & 0
\end{pmatrix}\ , &
\epsilon^{(l)}_{\mu\nu}&=
\begin{pmatrix} 
0 & 0 & 0 & 0 \\
0 & 0 & 0 & 0 \\
0 & 0 & 0 & 0 \\
0 & 0 & 0 & 1
\end{pmatrix}\ .
\end{align}
In more compact form,  the linear perturbation of tetrad $E_{\mu\nu}$,  traveling in the $+\hat{z}$ direction and keeping  $\mathbf{k}$ fixed,  may be expressed as 
\begin{multline} 
E_{\mu\nu}\left(t,z\right)=\sqrt{2}\left[\hat{\epsilon}^{(+)}\left(k_{z}\right)\epsilon^{(+)}_{\mu\nu}+\hat{\epsilon}^{(\times)}\left(k_{z}\right)\epsilon^{(\times)}_{\mu\nu}\right]e^{i\omega_{1}\left(t-z\right)}\\
+\hat{\epsilon}^{\left(s_{2}\right)}_{\mu\nu}\left(k_{z}\right)e^{i\left(\omega_{2}t-k_{z}z\right)}+\hat{\epsilon}^{\left(s_{3}\right)}_{\mu\nu}\left(k_{z}\right)e^{i\left(\omega_{3}t-k_{z}z\right)}+c.c.\ ,
\end{multline}
where $\hat{\epsilon}^{\left(s_{m}\right)}_{\mu\nu}$ is the polarization tensor associated to the scalar mode with $m\in\left\{2,3\right\}$
\begin{equation}
\hat{\epsilon}^{\left(s_{m}\right)}_{\mu\nu}\left(k_{z}\right)=\Biggl[\left(\frac{1}{2}+\frac{\omega^{2}_{m}}{k_{m}^{2}}\right)\epsilon_{\mu\nu}^{(TT)}-\frac{\sqrt{2}\omega_{m}k_{z}}{k_{m}^{2}}\epsilon_{\mu\nu}^{(TS)}\\
-\frac{1}{\sqrt{2}}\epsilon_{\mu\nu}^{(b)}+\left(-\frac{1}{2}+\frac{k_{z}^{2}}{k_{m}^{2}}\right)\epsilon_{\mu\nu}^{(l)}\Biggr]\frac{\hat{B}_{m}\left(k_{z}\right)}{3}\ .
\end{equation}

The two d.o.f. $\hat{\epsilon}^{\left(+\right)}$ and $\hat{\epsilon}^{\left(\times\right)}$ produce the two tensor modes associated to the polarization tensors $\epsilon_{\mu\nu}^{\left(+\right)}$ and $\epsilon_{\mu\nu}^{\left(\times\right)}$ while the two d.o.f. $\hat{B}_{2}$ and $\hat{B}_{3}$ produce two scalar modes associated to the polarization tensors $\hat{\epsilon}^{\left(s_{2}\right)}_{\mu\nu}$ and $\hat{\epsilon}^{\left(s_{3}\right)}_{\mu\nu}$. 
Indeed the polarization tensor $\hat{\epsilon}^{\left(s\right)}_{\mu\nu}$, restricted to  the spatial components $\hat{\epsilon}^{\left(s\right)}_{i,j}$,  is provided by 
\begin{equation}
\hat{\epsilon}^{\left(s_{m}\right)}_{i,j}=-\frac{1}{3\sqrt{2}}\hat{B}_{m}\left(k_{z}\right)\epsilon_{i,j}^{(b)}+\frac{1}{3}\left(-\frac{1}{2}+\frac{k_{z}^{2}}{k_{m}^{2}}\right)\hat{B}_{m}\left(k_{z}\right)\epsilon_{i,j}^{(l)}\ ,
\end{equation}
where $(i,j)$ range over $(1,2,3)$. Hence, the scalar mode is a combination of the longitudinal scalar mode and the transverse breathing scalar mode. 

More explicitly, for a massless plane wave $E_{\mu\nu}^{(k_{1})}$,  Eq.\eqref{eqdevgeolinear} gives
\begin{equation}
\begin{cases}

\ddot x(t)=\omega_{1}^{2}\left[\hat{\epsilon}^{\left(+\right)}\left(\omega_{1}\right)x+\hat{\epsilon}^{\left(\times\right)}\left(\omega_{1}\right)y\right]e^{i\omega_{1}\left(t-z\right)}+c.c. \\ \\
\ddot y(t)=\omega_{1}^{2}\left[\hat{\epsilon}^{\left(\times\right)}\left(\omega_{1}\right)x-\hat{\epsilon}^{\left(+\right)}\left(\omega_{1}\right)y\right]e^{i\omega_{1}\left(t-z\right)}+c.c.\\ \\
\ddot z(t)=0
\end{cases}\ ,
\end{equation}
where we obtain the two standard GW polarizations of  General Relativity, the purely transverse $+$ and  $ \times $ polarizations and the two-helicity massless tensor modes. 

Instead, for the $m$-th massive plane wave, $E_{\mu\nu}^{(k_{m})}$, where  $m=2,3$, with $k_{m}^{2}=M^{2}_{m}=\omega^{2}_{m}-k^{2}_{z}$ equal to the square of the $m$-th mass of the scalar field,   Eqs.\eqref{eqdevgeolinear} give
\begin{equation}
\begin{cases}
\ddot x(t)=-\frac{1}{6}\omega^{2}_{m}\hat{B}_{m}\left(k_{z}\right)x e^{i\left(\omega_{m}t-k_{z}z\right)}+c.c.\\ \\
\ddot y(t)=-\frac{1}{6}\omega^{2}_{m}\hat{B}_{m}\left(k_{z}\right)y e^{i\left(\omega_{m}t-k_{z}z\right)}+c.c.\\ \\
\ddot z(t)=-\frac{1}{6}M^{2}_{m}\hat{B}_{m}\left(k_{z}\right)ze^{i\left(\omega_{m}t-k_{z}z\right)}+c.c.
\end{cases}\ .
\end{equation}
This system of equations can be integrated assuming  small $E_{\mu\nu}\left(t,z\right)$  and, hence, we have 
\begin{equation}\label{soluzionseqgeod}
\begin{cases}
x(t)=x(0)+\frac{1}{6}\hat{B}_{m}\left(k_{z}\right)x(0)e^{i\left(\omega_{m}t-k_{z}z\right)}+c.c.\\ \\
y(t)=y(0)+\frac{1}{6}\hat{B}_{m}\left(k_{z}\right)y(0)e^{i\left(\omega_{m}t-k_{z}z\right)}+c.c.\\ \\
z(t)=z(0)+\frac{1}{6\omega^{2}_{m}}M_{m}^{2}\hat{B}_{m}\left(k_{z}\right)z(0)e^{i\left(\omega_{m}t-k_{z}z\right)}+c.c.
\end{cases}\ ,
\end{equation}
where we obtain  two further mixed massive scalar modes, zero-helicity, partially longitudinal and partially transverse breathing, that is each one with the same mixed scalar polarization. To visualize GW polarizations, we use  the geodesic deviations when a $m$-th   plane GW of frequency $\omega_{m}$ strikes a sphere of freely falling particles of radius $r=\sqrt{x^2(0)+y^{2}(0)+z^{2}(0)}$. The displacement of a given particle from the center of the ring $\vec{\chi}$ is given by the solution of the geodesic deviation equation \eqref{soluzionseqgeod}. The sphere will be distorted into an ellipsoid described by
\begin{equation}
\left(\frac{x}{\rho_{1m}(t)}\right)^{2}+\left(\frac{y}{\rho_{1m}(t)}\right)^{2}+\left(\frac{z}{\rho_{2m}(t)}\right)^{2}=r^{2}\ ,
\end{equation}
where both $\rho_{1m}(t)=1+\frac{1}{3}\hat{B}_{m}\left(k_{z}\right)\cos\left(\omega_{m}t-k_{z}z\right)$ and $\rho_{2m}(t)=1+\frac{M_{m}^{2}}{3\omega_{m}^{2}}\hat{B}_{m}\left(k_{z}\right)\cos\left(\omega_{m}t-k_{z}z\right)$ vary between their maximum and minimum value. Each swinging ellipsoid represents an additional scalar mode with zero-helicity which is partly longitudinal and partly
transverse \cite{PW}. The d.o.f. of the sixth-order teleparallel equivalent gravity are four: two of these, $\hat{\epsilon}^{\left(+\right)}$ and $\hat{\epsilon}^{\left(\times\right)}$, generate the $+$ and $\times$ tensor modes and the two d.o.f  $\hat{B}_{2}$ and $\hat{B}_{3}$ generate two mixed longitudinal-transverse scalar modes. Finally our theory of gravity shows three polarizations, two tensor and one mixed scalar polarizations, and four modes, two tensor  and two mixed scalar modes for each frequency $\omega_{2}$ and $\omega_{3}$. \\

The same results can be  obtained adopting the Newman-Penrose (NP) formalism for massless and massive waves  propagating along either null or nearly non-null geodesics \cite{CMW, NP}. To determine the {\it little group}\footnote{This is the subgroup of Lorentz transformations leaving the wave vector $\mathbf{k}$ invariant.} $E\left(2\right)$ classification of massive GWs\footnote{It is worth noticing that   this classification is strictly valid only for massless waves.}, we linearize the vacuum field equations in the limit of nearly massless plane waves (observer far from the source). We introduce a quasinormal local null tetrad basis $\left\{\hat{e}_{a}\right\}=\left(k, l, m, \bar{m}\right)$ as 
\begin{align}
k&=\frac{1}{\sqrt{2}}\left(\partial_{t}+\partial_{z}\right)\ ,
& l&=\frac{1}{\sqrt{2}}\left(\partial_{t}-\partial_{z}\right)\ ,\\
m&=\frac{1}{\sqrt{2}}\left(\partial_{x}+i\partial_{y}\right)\ ,
& \bar{m}&=\frac{1}{\sqrt{2}}\left(\partial_{x}-i\partial_{y}\right)\ ,
\end{align}
which satisfies the relations 
\begin{gather}
k\cdot l=-m\cdot\bar{m}=1\ , \nonumber\\
k\cdot k=l\cdot l=m\cdot m=\bar{m}\cdot\bar{m}=0\ ,\\
k\cdot m=k\cdot\bar{m}=l\cdot m=l\cdot\bar{m}=0\ . \nonumber
\end{gather}
Knowing that 
\begin{align}
\hat{e}_{a}=&e_{a}^{\mu}\partial_{\mu} & \theta^{a}=&\theta^{a}_{\mu}dx^{\mu}\ ,\\
g_{\mu\nu}=&\tilde{\eta}_{ab}\theta^{a}_{\mu}\theta^{b}_{\nu} & \tilde{\eta}_{ab}=&e_{a}^{\mu}e_{b}^{\nu}g_{\mu\nu}\ ,
\end{align}
with metric $\tilde{\eta}_{ab}$
\begin{equation}
\tilde{\eta}_{ab}=\tilde{\eta}^{ab}=
\begin{pmatrix} 
0 & 1 & 0 & 0 \\
1 & 0 & 0 & 0 \\
0 & 0 & 0 & -1 \\
0 & 0 & -1 & 0
\end{pmatrix}\ ,
\end{equation}
it is
\begin{equation}
g^{\mu\nu}=2k^{(\mu}l^{\nu)}-2m^{(\mu}\bar{m}^{\nu)}\ ,
\end{equation}
where $\left\{\theta^{a}\right\}$ is the dual tetrad of $\left\{\hat{e}_{a}\right\}$.
Now we express the fifteen NP scalars in terms of the four-dimensional Weyl tensor $C_{\mu\nu\rho\sigma}$ in a null tetrad basis defined as
\begin{equation}
C_{\mu\nu\rho\sigma}=R_{\mu\nu\rho\sigma}-2g_{[\mu|[\rho}R_{\sigma]|\nu]}+\frac{1}{3}g_{\mu[\rho}g_{\sigma]\nu}R\ .
\end{equation}
The five complex Weyl-NP scalars, classified with spin weight $s$, from the Weyl tensor expressed  in a null tetrad basis,  are
\begin{eqnarray}
&s=+2  & \Psi_{0}\equiv C_{kmkm}\ ,\nonumber\\
&s=+1  & \Psi_{1}\equiv C_{klkm}\ ,\nonumber\\
&s=0  & \Psi_{2}\equiv C_{km\bar{m}l}\ ,\\
&s=-1  & \Psi_{3}\equiv C_{kl\bar{m}l}\ ,\nonumber\\
&s=-2  & \Psi_{4}\equiv C_{\bar{m}l\bar{m}l}\ ,\nonumber
\end{eqnarray}
while the ten Ricci-NP scalars, classified with spin weight $s$, from the Ricci tensor expressed in a null tetrad basis, are
\begin{eqnarray*}
&s=2 & \Phi_{02}\equiv-\frac{1}{2}R_{mm\ ,}\\
&s=1 & \begin{cases}
\Phi_{01}\equiv-\frac{1}{2}R_{km}\\
\Phi_{12}\equiv-\frac{1}{2}R_{lm}\\
\end{cases}\ ,\\
&s=0 & 
\begin{cases}
\Phi_{00}\equiv-\frac{1}{2}R_{kk}\\
\Phi_{11}\equiv-\frac{1}{4}\left(R_{kl}+R_{m\bar{m}}\right)\\
\Phi_{22}\equiv-R_{ll}
\end{cases}\ ,\\
&s=-1 & 
\begin{cases}
\Phi_{10}\equiv-\frac{1}{2}R_{k\bar{m}}=\Phi_{01}^{*}\\
\Phi_{21}\equiv-\frac{1}{2}R_{l\bar{m}}=\Phi_{12}^{*}\\
\end{cases}\ ,\\
&s=-2 & \Phi_{20}\equiv-\frac{1}{2}R_{m\bar{m}}=\Phi_{02}^{*}\ ,\nonumber\\
& & \Lambda=\frac{R}{24}\ .
\end{eqnarray*} 
However, for  plane waves, they become 
\begin{align}
\Psi_{0}&=\Psi_{1}=0\ ,\\
\Psi_{2}&=\frac{1}{6}R_{lklk}\ ,\\
\Psi_{3}&=-\frac{1}{2}R_{lkl\bar{m}}\ ,\\
\Psi_{4}&=-R_{l\bar{m}l\bar{m}}\ ,\\
\Phi_{00}&=\Phi_{01}=\Phi_{10}=\Phi_{02}=\Phi_{20}=0\ ,\\
\Phi_{22}&=-R_{lml\bar{m}}\ ,\\
\Phi_{11}&=\frac{3}{2}\Psi_{2}\ ,\\
\Phi_{12}&=\bar{\Phi}_{21}=\bar{\Psi}_{3}\ ,
\end{align}
where we can choose the set $\left\{\Psi_{2}(u),\Psi_{3}(u),\Psi_{4}(u), \Phi_{22}(u)\right\}$ as  independent amplitudes to describe GWs propagating in the $+$z direction with the retarded time $u=t-z$.
Taking into account that the spatial components of driving-force matrix $S(t)$ of an ideal detector are the electric components  of the Riemann tensor $R_{i0j0}$ that is \cite{HKL, BEMI, AMdeA, dePMM, WAG, FSGS, YGSH}
\begin{equation}
S_{ij}(t)=R_{i0j0}\left(\tilde{u}\right)\ ,
\end{equation}
we express the response matrix $S(t)$ in terms of the six new basis polarization matrices $W_{A}(\mathbf{e}_{z})$ belonging to  to the wave direction $\mathbf{k}=\mathbf{e}_{z}$ with $A$ ranging over $\{1,2,3,4,5,6\}$, that is \cite{ELL,ELLWW}
\begin{equation}
S\left(t\right)=\sum_{A}p_{A}\left(\mathbf{e}_{z},t\right)W_{A}\left(\mathbf{e}_{z}\right)\ ,
\end{equation}
where 
\begin{align}
W_{1}\left(\mathbf{e}_{z}\right)=&-6\begin{pmatrix} 
0 & 0 & 0 \\
0 & 0 & 0 \\
0 & 0 & 1
\end{pmatrix}\ , & W_{2}\left(\mathbf{e}_{z}\right)=&-2\begin{pmatrix} 
0 & 0 & 1 \\
0 & 0 & 0 \\
1 & 0 & 0
\end{pmatrix}\ ,\nonumber\\
W_{3}\left(\mathbf{e}_{z}\right)=&2\begin{pmatrix} 
0 & 0 & 0 \\
0 & 0 & 1 \\
0 & 1 & 0
\end{pmatrix}\ , & W_{4}\left(\mathbf{e}_{z}\right)=&-\frac{1}{2}\begin{pmatrix} 
1 & 0 & 0 \\
0 & -1 & 0 \\
0 & 0 & 0
\end{pmatrix}\ ,\nonumber\\
W_{5}\left(\mathbf{e}_{z}\right)=&\frac{1}{2}\begin{pmatrix} 
0 & 1 & 0 \\
1 & 0 & 0 \\
0 & 0 & 0
\end{pmatrix}\ , & W_{6}\left(\mathbf{e}_{z}\right)=&-\frac{1}{2}\begin{pmatrix} 
1 & 0 & 0 \\
0 & 1 & 0 \\
0 & 0 & 0
\end{pmatrix}\ .
\end{align}
The six polarization amplitudes $p_{A}\left(\mathbf{e}_{z},t\right)$  can be expressed in terms of the six NP scalars, each with their own helicity as \cite{MN, WILL, AMA, MY, CCL, BCLN}
\begin{align}
p_{1}^{\left(l\right)}&\equiv\Psi_{2} & s&=0\ ,\\
p_{2}^{\left(x\right)}&\equiv\operatorname{Re}\Psi_{3} & s&=1\ ,\\
p_{3}^{\left(y\right)}&\equiv\operatorname{Im}\Psi_{3} & s&=-1\ ,\\
p_{4}^{\left(+\right)}&\equiv\operatorname{Re}\Psi_{4} & s&=2\ ,\\
p_{5}^{\left(\times\right)}&\equiv\operatorname{Im}\Psi_{4} & s&=-2\ ,\\
p_{6}^{\left(b\right)}&\equiv\Phi_{22} & s&=0\ .\\
\end{align}
The six polarizations modes are: the longitudinal mode $p_{1}^{\left(l\right)}$, the vector-$x$ mode $p_{2}^{\left(x\right)}$, the vector-$y$ mode $p_{3}^{\left(y\right)}$, the $+$ mode $p_{4}^{\left(+\right)}$, the $\times$ mode $p_{5}^{\left(\times\right)}$ and the breathing mode $p_{6}^{\left(b\right)}$.
Hence $\Psi_{2}$ and $\phi_{22}$ are both scalar modes, respectively purely transverse and purely longitudinal, the complex $\Psi_{3}$ are vector modes mixed and the complex $\Psi_{4}$ are the tensor modes purely transverse.\\

We conclude that the  GWs of  sixth-order teleparallel equivalent  gravity are composed by both massless plane waves $k_{1}^{2}=0$, moving along null-geodesics, and massive plane waves $k_{2}^{2},k_{3}^{2}\neq 0$, moving along non null-geodesics. In the hypothesis of slightly massive waves, we can use the NP-formalism for weak, plane, nearly null GWs \cite{CMW}. They have a wave 4-vector $k_{m}^{\mu}$, normal to surfaces of constant retarded time $\tilde{u}$, that is  
\begin{equation}
(k_{m})_{\mu}=-\tilde{u}_{,\mu}\ .
\end{equation} 
The GW propagation speed  $\vert\mathbf{v}_{g}\vert$ is less than $c$ for massive waves and  $c$ for massless waves 
\begin{equation}
\mathbf{v}_{g}=\frac{d\omega(\mathbf{k})}{d\mathbf{k}}=\frac{\mathbf{k}}{\omega(\mathbf{k})}\qquad\vert\mathbf{v}_{g}\vert\leq c\ .
\end{equation}
One can define a parameter  $\epsilon$ as the speed difference  between electromagnetic waves  and GWs in a local proper reference frame, that is 
\begin{equation}
\epsilon=\left(\frac{c}{v_{g}}\right)^2-1\ .
\end{equation}
in this local observer frame,  we may expand $k_{m}^{\mu}$ in terms of the  null-tetrad local basis $\left(k, l, m, \bar{m}\right)$, that is 
\begin{equation}
k_{m}^{\mu}=k^{\mu}\left(1+\epsilon_{k}\right)+\epsilon_{l}l^{\mu}+\epsilon_{m}m^{\mu}+\bar{\epsilon}_{m}\bar{m}^{\mu}\ ,
\end{equation}
where $\left\{\epsilon_{k},\epsilon_{l},\epsilon_{m}\right\}\sim \mathcal{O}\left(\epsilon\right)$ and $\epsilon$ vanishes for null wave because $\vert\mathbf{v}_{g}\vert=c$. The only NP quantities which are not $\mathcal{O}\left(\epsilon_{n}\right)$ are the four scalar $\left\{\Psi_{2}, \Psi_{3}, \Psi_{4}, \Phi_{22}\right\}$, with $\Psi_{3}$ and  $\Psi_{4}$ complex, that is
\begin{align}
\Psi_{2}=&-\frac{1}{6}R_{lklk}+\mathcal{O}\left(\epsilon_{l}R\right)\ ,\\
\Psi_{3}=&-\frac{1}{2}R_{lkl\bar{m}}+\mathcal{O}\left(\epsilon_{l}R\right)\ ,\\
\Psi_{4}=&-R_{l\bar{m}l\bar{m}}+\mathcal{O}\left(\epsilon_{l}R\right)\ ,\\
\Phi_{22}=&-R_{lml\bar{m}}+\mathcal{O}\left(\epsilon_{l}R\right)\ .
\end{align}
The linearized harmonic gauge in the perturbation $E_{\mu\nu}$ for our plane wave propagating in the $+$z direction becomes  
\begin{equation}
\begin{cases}
\partial_{0}E^{0}_{\phantom{0}0}+\partial_{3}E^{3}_{\phantom{0}0}=\frac{1}{2}\partial_{0}E\\
\partial_{0}E^{0}_{\phantom{0}1}+\partial_{3}E^{3}_{\phantom{0}1}=0\\
\partial_{0}E^{0}_{\phantom{0}2}+\partial_{3}E^{3}_{\phantom{0}2}=0\\
\partial_{0}E^{0}_{\phantom{0}3}+\partial_{3}E^{3}_{\phantom{0}3}=\frac{1}{2}\partial_{3}E
\end{cases}\ ,
\end{equation}
where $E$ is the trace of $E_{\mu\nu}$. After simple algebraic and differential calculations, we have 
\begin{equation}
\begin{cases}
\partial_{0}E_{01}=\partial_{3}E_{31}\\
\partial_{0}E_{02}=\partial_{3}E_{32}\\
\left(\partial_{00}-\partial_{33}\right)\left(E_{00}+E_{33}\right)=-\left(\partial_{00}+\partial_{33}\right)\left(E_{11}+E_{22}\right)\\
\left(\partial_{00}-\partial_{33}\right)E_{30}=-\partial_{03}\left(E_{11}+E_{22}\right)
\end{cases}\ ,
\end{equation}
which, after  derivations, give  
\begin{equation}
\begin{cases}
E_{01}=-\frac{k_{z}}{\omega}E_{31}\\
E_{02}=-\frac{k_{z}}{\omega}E_{32}\\
E_{11}+E_{22}=-\frac{\omega^{2}-k_{z}^{2}}{\omega^{2}+k_{z}^{2}}\left(E_{00}+E_{33}\right)\\
E_{30}=\frac{\omega k_{z}}{\omega^{2}-k_{z}^{2}}\left(E_{11}+E_{22}\right)
\end{cases}\ .
\end{equation}
Then the four NP-scalars become
\begin{align}
\Psi_{2}&=\frac{1}{6}\ddot{E}_{kk}+\mathcal{O}\left(\epsilon_{l}R\right)=-\frac{1}{6}\left[\frac{\omega^{2}-k_{z}^{2}}{\omega^{2}+k_{z}^{2}}\left(\omega^{2}E_{33}-k_{z}^{2}E_{00}\right)\right]+\mathcal{O}\left(\epsilon_{l}R\right)\ ,\\
\Psi_{3}&=\frac{1}{2}\ddot{E}_{k\bar{m}}+\mathcal{O}\left(\epsilon_{l}R\right)=-\frac{1}{2}\left(\omega^{2}-k_{z}^{2}\right)E_{13}+\frac{1}{2}i\left(\omega^{2}-k_{z}^{2}\right)E_{23}+\mathcal{O}\left(\epsilon_{l}R\right)\ ,\\
\Psi_{4}&=\ddot{E}_{\bar{m\bar{m}}}+\mathcal{O}\left(\epsilon_{l}R\right)=\left(\frac{\omega^{2}-k_{z}^{2}}{\omega^{2}+k_{z}^{2}}\right)\omega^{2}\left(E_{00}+E_{33}\right)+2\omega^{2}E_{22}+2i\omega^{2}E_{12}+\mathcal{O}\left(\epsilon_{l}R\right)\ ,\\
\Phi_{22}&=\ddot{E}_{m\bar{m}}+\mathcal{O}\left(\epsilon_{l}R\right)=\left(\frac{\omega^{2}-k_{z}^{2}}{\omega^{2}+k_{z}^{2}}\right)\omega^{2}\left(E_{00}+E_{33}\right)+\mathcal{O}\left(\epsilon_{l}R\right)\ ,
\end{align}
where $\ddot{E}=E_{,\tilde{u}\tilde{u}}$ and $\tilde{u}=t-v_{g}z$.\\

Hence by Eqs.\eqref{FirstOrderTetradmassless} and \eqref{FirstOrderTetradmassive}, we get, for massless modes $\omega_{1}$ associated to the plane wave $E_{\mu\nu}^{k_{1}}$, the following NP-quantities 
\begin{equation}
\Psi_{2}=\Psi_{3}=\Phi_{22}=0\ ,
\end{equation}
\begin{equation}
0\neq\Psi_{4}=-2\omega_{1}^{2}\hat{\epsilon}^{\left(+\right)}\left(\omega_{1}\right)+2i\omega_{1}^{2}\hat{\epsilon}^{\left(\times\right)}\left(\omega_{1}\right)\left(e^{i\omega_{1}\left(t-z\right)}+c.c.\right)\ ,
\end{equation}
and amplitudes
\begin{align}\label{masspolarampl}
p_{1}^{\left(l\right)}\left(\mathbf{e}_{z},t\right)&=p_{2}^{\left(x\right)}\left(\mathbf{e}_{z},t\right)=p_{3}^{\left(y\right)}\left(\mathbf{e}_{z},t\right)=p_{6}^{\left(b\right)}\left(\mathbf{e}_{z},t\right)=0\ ,\nonumber\\
p_{4}^{\left(+\right)}\left(\mathbf{e}_{z},t\right)&=-2\omega_{1}^{2}\hat{\epsilon}^{\left(+\right)}\left(\omega_{1}\right)e^{i\omega_{1}\left(t-z\right)}+c.c.\ ,\nonumber\\
p_{5}^{\left(\times\right)}\left(\mathbf{e}_{z},t\right)&=2\omega_{1}^{2}\hat{\epsilon}^{\left(\times\right)}\left(\omega_{1}\right)e^{i\omega_{1}\left(t-z\right)}+c.c.\ ,
\end{align}
that is, for frequency $\omega_{1}$, we obtain the two standard $+$ and $\times$ transverse modes with  helicity  2 and the $E\left(2\right)$ classification is $N_{2}$. More precisely the $+$ mode is generated by the d.o.f. $\hat{\epsilon}^{\left(+\right)}\left(\omega_{1}\right)$ while the $\times$ mode is generated by the d.o.f.  $\hat{\epsilon}^{\left(\times\right)}\left(\omega_{1}\right)$.\\

On the other hand,  by Eqs.\eqref{FirstOrderTetradmassless} and \eqref{FirstOrderTetradmassive} for massive modes $\omega_{2}$ and $\omega_{3}$ associated to the plane wave $E_{\mu\nu}^{k_{m}}$ with $m$ ranging over $\left\{2,3\right\}$, we get the following NP-quantities 
\begin{equation}
\Psi_{3}=\Psi_{4}= 0\ ,
\end{equation}
\begin{equation}
0\neq\Psi_{2}=\frac{1}{36}\left(\omega_{m}^{2}-k_{z}^{2}\right)\hat{B}_{m}\left(k_{z}\right)e^{i\left(\omega_{m}t-k_{z}z\right)}+\mathcal{O}\left(\epsilon_{l}R\right)+c.c.\ ,
\end{equation}
\begin{equation}
0\neq\Phi_{22}=\frac{\omega_{m}^{2}}{3}\hat{B}_{m}\left(k_{z}\right)e^{i\left(\omega_{m}t-k_{z}z\right)}+\mathcal{O}\left(\epsilon_{l}R\right)+c.c.\ ,
\end{equation}
and amplitudes
\begin{align}\label{masspolarampl}
p_{1}^{\left(l\right)}\left(\tilde{\mathbf{e}}_{z},t\right)&=\frac{1}{36}\left(\omega_{m}^{2}-k_{z}^{2}\right)\hat{B}_{m}\left(k_{z}\right)e^{i\left(\omega_{m}t-k_{z}z\right)}+\mathcal{O}\left(\epsilon_{l}R\right)+c.c.\ ,\nonumber\\
p_{2}^{\left(x\right)}\left(\tilde{\mathbf{e}}_{z},t\right)&=p_{3}^{\left(y\right)}\left(\mathbf{e}_{z},t\right)=p_{4}^{\left(+\right)}\left(\mathbf{e}_{z},t\right)=p_{5}^{\left(\times\right)}\left(\mathbf{e}_{z},t\right)=0\ ,\nonumber\\
p_{6}^{\left(b\right)}\left(\tilde{\mathbf{e}}_{z},t\right)&=\frac{\omega_{m}^{2}}{3}\hat{B}_{m}\left(k_{z}\right)e^{i\left(\omega_{m}t-k_{z}z\right)}+\mathcal{O}\left(\epsilon_{l}R\right)+c.c.\ ,
\end{align}
that is for each frequency $\omega_{2}$ and $\omega_{3}$,  we obtain one mixed longitudinal-transverse scalar mode with helicity 0 and the $E\left(2\right)$ classification is $II_{6}$. More specifically both scalars $\Psi_{2}$ and $\Phi_{22}$ are driven by one d.o.f.  $B_{m}$ for each frequency $\omega_{m}$. Specifically, this fact means that both longitudinal and breathing modes, related to each $\omega_{m}$, show themselves always coupled in a single mixed scalar mode. 
In conclusion,  we obtained three polarizations, two tensor polarizations and one mixed scalar polarization, and four modes, two tensor modes and two scalar modes. This result is   because the two mixed scalar modes, associated with each frequency,  exhibit the same polarization. If we go up with the order of the theory equivalent to polynomial curvature Lagrangian $f\left(R, \Box R,\dots, \Box^{k}R\right)$, for every two orders, a single mixed longitudinal-transverse scalar mode has to be  added as shown in  Table \ref{pol ed eli}.  

\section{Higher-Order Teleparallel Gravity not equivalent to Higher-Order Curvature Gravity}
Finally let us  investigate  a generic teleparallel Lagrangian of order $n$ which does not contain the boundary term $B$. To get  the field equations of any order $n$,  we have to  introduce the linear differential terms $\Box^{k} T$, in analogy with the Lagrangian $L_{\Box^{k}R}=\sqrt{-g}\left(R+\sum_{k=0}^{p}a_{k}R\Box^{k} R\right)$, that is
\begin{equation}
L_{\Box^{k}T}^{neq}=\frac{e}{2\kappa^{2}}\left(T+a_{0}T^{2}+\sum_{k=1}^{p} a_{k}T\Box^{k}T\right)\ .
\end{equation}
By varying the  Lagrangian $L_{\Box^{k}T}^{neq}$ plus the material Lagrangian term $L_{m}$ with respect to a tetrad basis $e^{a}_{\ \rho}$, we obtain the  $(2p + 2)$-order equations:
\begin{equation}\label{FEHOTG}
\boxed{
\begin{split}
\frac{4}{e}\partial_{\sigma}\left[e\left(1+2a_{0}T+2\sum_{k=1}^{p}a_{k}\Box^{k}T\right)S_{a}^{\phantom{a}\rho\sigma}\right]-4\left(1+2a_{0}T+2\sum_{k=1}^{p}a_{k}\Box^{k}T\right)T^{\mu}_{\ \nu a}S_{\mu}^{\phantom{\mu}\nu\rho}\\
+\left(T+a_{0}T^{2}+\sum_{k=1}^{p} a_{k}T\Box^{k}T\right)e_{a}^{\phantom{a}\rho}
-e_{a}^{\phantom{a}\rho}\sum_{k=1}^{p}a_{k}\sum_{h=1}^{k}\Box^{h-1}T\left(\nabla^{\rho}\nabla_{\nu}+\nabla_{\nu}\nabla^{\rho}\right)\Box^{k-h}T\\
+\sum_{k=1}^{p}a_{k}\sum_{h=1}^{k}\Box^{h-1}T\partial_{\eta}\Box^{k-h}T\left(e_{a}^{\phantom{a}\eta}e_{b}^{\phantom{b}\rho}\partial^{\nu}e^{b}_{\phantom{b}\nu}+T_{a}^{\phantom{a}\eta\rho}-e_{a}^{\eta}T^{\rho}-g^{\eta\rho}T_{a}\right)\\
-\sum_{k=1}^{p}a_{k}\sum_{h=1}^{k}\partial_{\sigma}\left[\left(e_{a}^{\phantom{a}\rho}g^{\eta\sigma}-e_{a}^{\phantom{a}\sigma}g^{\eta\rho}-e_{a}^{\ \eta}g^{\sigma\rho}\right)\Box^{h-1}T\partial_{\eta}\Box^{k-h}T\right]=2\kappa^{2}\mathcal{T}^{\left(m\right)\phantom{a}\rho}_{\phantom{\left(m\right)}a}\ .
\end{split}
}
\end{equation}
Thus, we linearize field Eqs. \eqref{FEHOTG} keeping up to first order terms in $E^{a}_{\phantom{a}\mu}$. We get
\begin{equation}
2\partial_{\sigma}S_{\tau}^{\phantom{\tau}\rho\sigma\left(1\right)}=\kappa^{2}\mathcal{T}_{\tau}^{\phantom{\tau}\rho\left(0\right)}\ .
\end{equation}
Adopting the approximation \eqref{2DS1O},  one obtains 
\begin{equation}\label{LFEGLHOTG}
\Box\bar{E}^{\mu}_{\phantom{\mu}\rho}=-\kappa^{2}\mathcal{T}_{\tau}^{\phantom{\tau}\rho\left(0\right)}\ .
\end{equation}
Solving Eq.\eqref{LFEGLHOTG} in vacuum we get the solutions
\begin{equation}
E_{\rho\tau}\left(x\right)=\int \frac{d^{3}\mathbf{k}}{(2\pi)^{3/2}}\hat{C}_{\rho\tau}\left(\mathbf{k}\right)e^{ik^{\alpha}x_{\alpha}}+c.c.\ ,
\end{equation}
with $k^{2}=0$ which, for a wave propagating along the $+\hat{z}$ direction, becomes 
\begin{equation}
E_{\mu\nu}\left(t,z\right)=\int\frac{d^{3}\mathbf{k}}{(2\pi)^{3/2}}\sqrt{2}\left[\hat{\epsilon}^{(+)}\left(\mathbf{k}\right)\epsilon^{(+)}_{\mu\nu}+\hat{\epsilon}^{(\times)}\left(\mathbf{k}\right)\epsilon^{(\times)}_{\mu\nu}\right]e^{i\omega\left(t-z\right)}+c.c.\ ,.
\end{equation}
This results means that  we achieve the Einstein GWs with two transverse tensor polarizations and helicity 2. In other words, the absence of boundary term $B$ in the Lagrangian  generates only  the two standard $+$ and $\times$ polarizations. Further modes emerge if we  introduce the boundary term $B$ that excites extra polarizations. This result generalizes the one obtained in \cite{bamba} for $f(T)$ gravity.  In other words, higher-order teleparallel gravity without boundary terms is equivalent to General Relativity from the point of view of GWs.\\
A summary of polarizations and helicity states is reported in Table \ref{pol ed eli} for several extended teleparallel theories.
\begin{table}[h]
\begin{equation*}
\footnotesize{
\begin{array}{ccccccccc}
\toprule
\text{Lagrangian}  & \text{Order} & \text{Frequency} & \text{Polarization} & \text{Type} & \text{Modes} & \text{State} & \text{Helicity} & \text{Mass}\\
&&&&&\text{d.o.f.} &&&\\
\midrule
L_{f(T)}  & 2\text{th} & \omega_{1} & 2 & \text{transverse tensor} & 2 & \epsilon_{\mu\nu}^{(+)},\epsilon_{\mu\nu}^{(\times)} & 2 &  0 \\
\midrule
L_{f(T,B)} & 4\text{th} & \omega_{1}, \omega_{2} & 3 & \text{transverse tensor}  & 3 & \epsilon_{\mu\nu}^{(+)},\epsilon_{\mu\nu}^{(\times)} & 2,0 & M_{2}\\
 & &  & & \text{mixed scalar} & & \epsilon_{\mu\nu}^{(s_{2})} & & \\
\midrule
L_{\Box T}^{eq} & 6\text{th} & \omega_{1}, \omega_{2} & 3 & \text{transverse tensor } & 4 & \epsilon_{\mu\nu}^{(+)},\epsilon_{\mu\nu}^{(\times)} & 2,0 & M_{3}\\
 & & \omega_{3} & & \text{mixed scalar} & & \epsilon_{\mu\nu}^{(s_{2})}, \epsilon_{\mu\nu}^{(s_{3})} & & \\
\midrule
L_{\Box^{2} T}^{eq} & 8\text{th} & \omega_{1}, \omega_{2} & 3 & \text{transverse tensor} & 5 &  \epsilon_{\mu\nu}^{(+)},\epsilon_{\mu\nu}^{(\times)} & 2,0 & M_{4}\\
& & \omega_{3},\omega_{4} & & \text{mixed scalar} & & \epsilon_{\mu\nu}^{(s_{2})}, \epsilon_{\mu\nu}^{(s_{3})} & & \\
&&&&&&\epsilon_{\mu\nu}^{(s_{4})}&&\\
\midrule
\vdots & \vdots & \vdots & \vdots &\vdots & \vdots & \vdots & \vdots\\
\midrule
L_{\Box^{p}T}^{eq} & 2(p+2)\text{th} & \omega_{1},\dots & 3 & \text{transverse tensor} & p+3 & \epsilon_{\mu\nu}^{(+)},\epsilon_{\mu\nu}^{(\times)} & 2, 0 & M_{p+2}\\
& & \dots ,\omega_{p+2} & & \text{mixed scalar} &  & \epsilon_{\mu\nu}^{(s_{2})},\dots  & &\\ 
&&&&&&\dots,\epsilon_{\mu\nu}^{(s_{p+2})}&&\\
\midrule
L_{\Box^{k}T}^{neq}& 2(p+1)\text{th} & \omega=q & 2 & \text{tensor} & 2 &\epsilon_{\mu\nu}^{(+)},\epsilon_{\mu\nu}^{(\times)}  & 2 & 0\\
\bottomrule
\end{array}
}
\end{equation*}
\caption{Classification of Gravitational Waves for Extended Teleparallel Theories of Gravity }
\label{pol ed eli}
\end{table}
\section{Conclusions}
Teleparallelism is an approach to  gravity  equivalent to General Relativity,  if torsion and curvature scalars are linear into the gravitational Lagrangian.  Considering  $f\left(R\right)$ and $f\left(T\right)$ theories, or, more generally,
 higher-order theories  such as $f\left(R, \Box R,\dots, \Box^{k}R\right)$ and $f(T, \Box T,$ $\dots, \Box^{k}T)$ the non-equivalence emerges   because the involved d.o.f.  are different. To describe the same  physics,  one  has to   take into account the boundary term $B$ with its derivatives.  The equivalence is  restored according to the relation \eqref{equivalence}.
 
  As shown above,  we derived the teleparallel Lagrangian equivalent to   $L_{\Box R}=-R+a_{0}R^{2}+a_{1}R\Box R$  considering the boundary term $B$ and its derivatives.  The linearized field equations, in the low energy regime, have been solved in vacuum adopting a distribution theory approach. Thus, we have derived the  GWs of  sixth-order equivalent teleparallel gravity and investigated their properties. GWs show three polarizations and four oscillation modes related to the four d.o.f. More in detail, in addition to the two tensor modes of General Relativity of frequency $\omega_{1}$, both  transversely polarized, massless, with  helicity 2 and  propagation speed $c$, we  obtained two further scalar modes associated with each frequency $\omega_{2}$ and $\omega_{3}$. Every individual scalar mode is partly transversally polarized and partly longitudinally polarized because governed by a single d.o.f., respectively $\hat{B}_{2}\left(k_{z}\right)$ and $\hat{B}_{3}\left(k_{z}\right)$: they form a single mixed transverse-longitudinal massive scalar mode with  helicity 0 and propagation speed  less than $c$. These results can be  both demonstrated adopting the  geodesic deviation equation  and the Newmann-Penrose formalism, generalized to slightly massive waves. It is important to point out that in every higher-order teleparallel theory, equivalent to the polynomial Lagrangian $f\left(R, \Box R,\dots, \Box^{k}R\right)$,  every two higher-orders excite a  mixed scalar mode.  On the other hand, considering theories like 
$f\left(T, \Box T,\dots, \Box^{k}T\right)$ is completely equivalent to General Relativity from the point of view of GWs.  We stress again the crucial relevance of the boundary term $B$ to improve the number of GW modes and polarizations.

 This means that  precise polarization measurements cannot only  invalidate or promote a given theory of gravity \cite{BCLN, GW17_1, GW17_2, GW18, lombriser, CCT2, CCT1} but also suggest if metric or teleparallel representation better represent the gravitational interaction. See also \cite{CCLS} for a discussion on this point.  

 \section*{Acknowledgements}
 SC is supported in part by the INFN sezione di Napoli, {\it iniziative specifiche} MOONLIGHT2 and QGSKY.

\appendix 
\section{Appendix: Differential Identities}\label{A}
Let $f$ and $g$ be two scalar functions and $A^{\mu}$ a controvariant vector, the following differential identities hold:
\begin{equation}
f\Box g=\nabla^{\nu}\left(f\nabla_{\nu}g-g\nabla_{\nu}f\right)+g\Box f\ ,
\end{equation}
\begin{equation}
\Box\left(fg\right)=f\Box g+g\Box f+2\nabla^{\mu}f\nabla_{\mu}g\ ,
\end{equation}
\begin{equation}
f\Box^{k}g=\nabla^{\nu}\left(\cdots\right)_{\nu}+g\Box^{k}f\ ,
\end{equation}
\begin{equation}
\partial_{\mu}\left(e A^{\mu}\right)=e\nabla_{\mu}A^{\mu}\ .
\end{equation}
\section{Appendix: Main Variations}\label{B}
Here a summary of main variations adopted along the text:
\begin{equation}
\delta e=e e_{a}^{\phantom{a}\rho}\delta e^{a}_{\phantom{a}\rho}=-e e^{a}_{\phantom{a}\rho}\delta e_{a}^{\phantom{a}\rho}\ ,
\end{equation}
\begin{equation}
\delta e_{a}^{\phantom{a}\rho}=-e_{a}^{\phantom{a}\sigma}e_{b}^{\phantom{a}\rho}\delta e^{b}_{\phantom{b}\sigma}\ ,
\end{equation}
\begin{equation}
\delta g^{\mu\nu}=\left(-g^{\nu\rho}e_{a}^{\phantom{a}\mu}-g^{\mu\rho}e_{a}^{\phantom{a}\nu}\right)\delta e^{a}_{\phantom{a}\rho}\ ,
\end{equation}
\begin{equation}
\delta T=-4 T^{\mu}_{\phantom{\mu}\nu a}S_{\mu}^{\phantom{\mu}\nu\rho}\delta e^{a}_{\ \rho}-4S_{a}^{\phantom{a}\rho\sigma}\delta\left(\partial_{\sigma} e^{a}_{\phantom{a}\rho}\right)\ ,
\end{equation}
\begin{equation}
\delta\Box T=\nabla_{\mu}\nabla_{\nu}T\delta g^{\mu\nu}+\Box\delta T-g^{\mu\nu}\left(\delta\tilde{\Gamma}^{\eta}_{\ \nu\mu}-\delta K^{\eta}_{\ \nu\mu}\right)\partial_{\eta}T\ ,
\end{equation}
\begin{equation}
\begin{split}
\delta \Box T=\left[-e_{a}^{\ \nu}\left(\nabla^{\rho}\nabla_{\nu}+\nabla_{\nu}\nabla^{\rho}\right)T+\left(e_{a}^{\ \eta}e_{b}^{\ \rho}\partial^{\nu}e^{b}_{\ \nu}+T_{a}^{\ \eta\rho}-e_{a}^{\ \eta}T^{\rho}-g^{\eta\rho}T_{a}\right)\partial_{\eta}T\right]\delta e^{a}_{\ \rho}\\
+\left(e_{a}^{\ \rho}g^{\eta\sigma}-e_{a}^{\ \sigma}g^{\eta\rho}-e_{a}^{\ \eta}g^{\rho\sigma}\right)\partial_{\eta}T\delta\left(\partial_{\sigma} e^{a}_{\ \rho}\right)+\Box\delta T\ ,
\end{split}
\end{equation}
\begin{equation}
T\delta\Box^{k}T=-2e_{a}^{\phantom{a}\nu}\sum_{h=1}^{k}\Box^{h-1}T\nabla^{(\rho}\nabla_{\nu)}\Box^{k-h}T\delta e^{a}_{\phantom{a}\rho}-e\sum_{h=1}^{k}\Box^{h-1}T\partial_{\eta}\Box^{k-h}Tg^{\mu\nu}\delta\stackrel{\circ}{\Gamma}{\hspace{-0.04in}}^{\eta}_{\phantom{\eta}\nu\mu}+e\Box^{k}T\delta T\ ,
\end{equation}
\begin{equation}
\delta T^{\nu\lambda}_{\phantom{\nu\lambda}\nu}=-\left(T^{\rho\lambda}_{\phantom{\rho\lambda}a}+e_{a}^{\phantom{a}\lambda}T^{\rho}+g^{\lambda\rho}T_{a}\right)\delta e^{a}_{\phantom{a}\rho}+\left(e_{a}^{\phantom{a}\rho}g^{\lambda\sigma}-e_{a}^{\phantom{a}\sigma}g^{\lambda\rho}\right)\delta\left(\partial_{\sigma} e^{a}_{\phantom{a}\rho}\right)\ ,
\end{equation}
\begin{equation}
\delta B = \delta \left[\frac{2}{e}\partial_{\sigma}\left(e T^{\nu\sigma}_{\phantom{\nu\sigma} \nu}\right)\right]=-\frac{B}{e}\delta e+\frac{2}{e}\partial_{\mu}\left(T^{\mu}\delta e+e\delta T^{\mu}\right)\ .
\end{equation}

\end{document}